\documentclass[structabstract]{aa}  
\usepackage{graphicx}
\usepackage{txfonts}

\newcommand{\2}{{~\sc ii}}

\newcommand{\mic}{{\,$\mu$m}}

\begin{document}
   \title{Oxygen-rich dust production in IC\,10}
   \titlerunning{Oxygen-rich dust production in IC\,10}

   %\subtitle{I. Overviewing the $\kappa$-mechanism}

   \author{V. Lebouteiller\inst{1,2}, G.\,C.\ Sloan\inst{2}, M.\,A.\,T.\ Groenewegen\inst{3}, 
          M.\ Matsuura\inst{4}, D.\ Riebel\inst{5,6}, D.\,G.\ Whelan\inst{7}, J.\ Bernard-Salas\inst{8}, P.\ Massey\inst{9} \and E.\ Bayet\inst{10}
          }
          \authorrunning{Lebouteiller et al.}

   \institute{
   $^1$\ Laboratoire AIM, CEA/DSM-CNRS-Universit\'e Paris Diderot DAPNIA/Service d'Astrophysique B\^at. 709, CEA-Saclay F-91191 Gif-sur-Yvette C\'edex, France\\  \email{vianney.lebouteiller@cea.fr}, \email{vianney@isc.astro.cornell.edu} \\
             $^2$\ Center for Radiophysics and Space Research, Cornell University, Space Sciences Building, Ithaca, NY 14853-6801, USA\\
             $^3$\ Koninklijke Sterrenwacht van Belgi\"e, Ringlaan 3, B--1180 Brussels, Belgium \\
             $^4$\ Astrophysics Group, Department of Physics and Astronomy, University College London, Gower Street, London WC1E 6BT, UK\\
             $^5$\ Department of Physics and Astronomy, The Johns Hopkins University, 3400 North Charles St. Baltimore, MD 21218, USA\\
             $^6$\ Department of Physics, United States Naval Academy, 572C Holloway Road, Annapolis, MD 21402, USA	\\
             $^7$\ Department of Astronomy, University of Virginia, P.O. Box 400325, Charlottesville, VA 22904, USA\\
             $^8$\ Institut dÕAstrophysique Spatiale, CNRS/Universit{\'e} Paris-Sud 11, 91405 Orsay, France\\
             $^9$\ Lowell Observatory, 1400 West Mars Hill Rd., Flagstaff, AZ 86001, USA\\
             $^{10}$\ Sub-Department of Astrophysics, University of Oxford, Denys Wilkinson Building, Keble Road, Oxford OX1 3RH, UK\\
             }

   \date{Received ; accepted }

% \abstract{}{}{}{}{} 
% 5 {} token are mandatory
 
  \abstract
   {We report the detection of oxygen-rich circumstellar envelopes in stars of the nearby ($700$\,kpc) starburst galaxy IC\,10. The star formation history and the chemical environment of this galaxy makes it an ideal target to observe dust production by high-mass stars in a low-metallicity environment.  }
   {The goal of this study is to identify oxygen-rich stars in IC\,10 and to constrain their nature between asymptotic giant branch stars (AGBs), red supergiants (RSGs), and other infrared bright sources. We examine the mass-loss rate of the stars and compare to results obtained for the Magellanic Clouds. Our objectives are to (1) assess whether RSGs can be significant dust producers in IC\,10, and (2), solve the discrepancy between the star formation history of IC\,10 and the relatively low number of RSGs detected in the optical.  }
   {We search for silicate dust in emission by using the spectral map observed with the Infrared Spectrograph on board the Spitzer Space Telescope. The optical ($UBVRI$) and infrared ($JHK$, Spitzer/IRAC and Spitzer/MIPS) photometry is used to assert the membership of the stars to IC\,10 and disentangle between AGBs and RSGs. Radiative models are used to infer mass-loss rates and stellar luminosities.}
   {The luminosity and colors of at least $9$ silicate emission sources are consistent with stars within IC\,10. Furthermore, the photometry of $2$ of these sources is consistent with RSGs. We derive dust mass-loss rates similar to the values found in the Magellanic Clouds. Accounting for the sample completeness, RSGs are not important contributors to the dust mass budget in IC\,10.   }
   {}


   \maketitle
%
%________________________________________________________________

\section{Introduction}

Principal contributors to the dust in the interstellar medium (ISM) are thought to be low-mass asymptotic giant branch (AGB) stars, supernovae (SNe), red supergiants (RSGs), and late-type WC Wolf-Rayet stars (e.g., Gehrz 1989). AGB stars are expected to contribute to significant dust production in the most metal-poor sources, because their evolution timescales are shorter in such environments. For metallicities as low as $1/200$\,Z\,$_\odot$, stars might take as short as 100\,Myr to evolve from the zero-age main sequence to the AGB (Ventura et al.\ 2002; Herwig 2004). Boyer et al.\ (2012) found that AGBs produce most of the dust from cool evolved stars in the metal-poor Small Magellanic Cloud (SMC, $0.2$\,Z$_\odot$), but other dust production sources such as growth from existing grains are necessary to explain the total dust mass budget.  
%While it was expected that at low-Z AGBs may not produce enough dust, studies by Sloan et al.\ (2012) show that C-rich stars seem to produce all the dust they need to drive outflows. 
Alternatively, it is largely debated whether cosmic dust abundance can be reconciled with SN dust (e.g., Li et al.\ 2008; Maiolino et al.\ 2004), as dust mass produced in SNe is largely uncertain ($10^{-3}$\,M$_\odot$ up to $0.5$\,M$_\odot$; e.g., Stanimirovic et al.\ 2005; Matsuura et al.\ 2011). 
With these considerations in mind, high-mass stars could be significant contributors to the ISM dust. In metal-poor galaxies, where late-type WCs are scarce, RSGs might be the dominant dust source (Massey et al.\ 2005; but see Boyer et al.\ 2012).  
%RSGs should therefore be considered as a potential source of stardust in the early Universe.

The detection of infrared excess due to mass loss in the circumstellar envelopes of evolved stars within Local Group galaxies is mostly limited to low-mass carbon-rich stars ($\sim1$\,M$_\odot$; e.g., Sloan et al.\ 2012 and references therein). This is due to the star formation history of our nearest neighbor galaxies, with the lack of a recent starburst episode that would result in pronounced populations of $-$ more massive $-$ oxygen-rich stars. From an observational point of view, probing O-rich dust production in Local Group galaxies is challenging as it requires observing the silicate emission bands in the mid-infrared range. The observation of such stars has been therefore limited only to the Milky Way and to the Magellanic Clouds with the space telescopes ISO (e.g., Trams et al.\ 1999a; 1999b) and Spitzer (e.g., Buchanan et al.\ 2009; Groenewegen et al.\ 2009; van Loon et al.\ 2010; Woods et al.\ 2011; Boyer et al.\ 2011). The great sensitivity of Spitzer along with matured data analysis techniques now makes it possible to observe O-rich dust spectra beyond the Magellanic Clouds and in starburst galaxies.  

The nearby dwarf starburst IC\,10 is an ideal target to detect O-rich stars in a metal-poor environment, such as what has been accomplished for the Magellanic Clouds. IC\,10 was discovered by Mayall (1935) and Hubble (1936); it is the nearest starburst galaxy known ($\sim$700\,kpc; Kim et al.\ 2009; Kennicutt et al.\ 1998; Borissova et al.\ 2000; Hunter 2001). Its size and mass are comparable to the SMC while its metallicity of $12+\log$(O/H)$\approx8.26$ ($1/2.7$\,Z$_\odot$ assuming the solar abundance from Asplund et al.\ 2009) lies between that of the Small and Large Magellanic Cloud (Garnett 1990; Lequeux et al.\ 1979; Richer et al.\ 2001; Skillman, Kennicut \& Hodge 1989). The starburst nature of IC\,10 was first revealed by the discovery of a large number WR stars by Massey et al.\ (1992). IC\,10 has experienced several episodes of extensive star formation, with the most recent ones a few 10s to 100s of Myr ago (Vacca et al.\ 2007). 
%These starburst episodes make this galaxy a unique case in underlying infrared stellar population from other metal-poor local dwarf galaxies, where the significant star formation was a few Giga years ago (). 
The presence of a widespread population of WR stars (Massey et al.\ 1992; Massey \&\ Homes 2002; Royer et al.\ 2001) suggests that IC\,10 starburst is also widespread and that RSGs are to be expected throughout the galaxy. At low metallicities, the number of RSGs should even dominate over the number of WR stars, as one expects
from evolutionary theory (Maeder et al.\ 1980) and as demonstrated observationally in the Local Group by Massey (2002, 2003). The number of spectroscopically confirmed IC\,10 WRs is $24$, and the actual number is believed to be many more (Massey \& Holmes 2002). From the relatively low metallicity of IC\,10, one would thus expect the population of RSGs to be about $50-100$ strong (see Figure 12 of Massey 2003), not inconsistent with the color-magnitude diagram (CMD) of this galaxy. 

After presenting the observations in Section\,\ref{sec:obs}, we derive a preliminary silicate emission map of IC\,10 from which we identify several O-rich candidates (Section\,\ref{sec:map}). We then cross-correlate our sample with optical and infrared catalogs, removing foreground stars in the process (Section\,\ref{sec:sample}). The final mid-infrared spectra (Section\,\ref{sec:spectra}) are used to derive mass-loss rates and discuss the stellar chemistry of the sources (Section\,\ref{sec:properties}).

%__________________________________________________________________

\section{Observations}\label{sec:obs}

IC\,10 was observed with the Infrared Spectrograph (IRS; Houck et al.\ 2004) onboard the Spitzer Space Telescope (Werner et al.\ 2004) on 2008 September 13 as part of GTO program 50318. Observations consisted in a sparse spectral map with the \textit{Short-Low} (SL) module, providing the wavelength coverage between $\approx5-14.5$\mic\ with a spectral resolution power between $60$ and $130$. The map is made of $58$ perpendicular steps and $8$ parallel steps, with 2 cycles of 14 seconds per exposure. The galaxy was not fully sampled spatially because of time constraints. A gap was deliberately introduced between every perpendicular scanning position, with the gap size precisely equal to the width of the \textit{Short-Low} aperture ($\approx3.7\arcsec$). The full width at half maximum (FWHM) of the point spread function is on the order of the aperture width ($\approx3.5\arcsec$ at $10$\mic), allowing the detection of light from outside the slit despite the gaps. The flux is calibrated by performing optimal extraction of the point-like sources and accounting for the slit throughput (Section\,\ref{sec:spectra}). In order to solve the incomplete spatial sampling for the preliminary analysis (Section\,\ref{sec:map}), gaps were interpolated using a cubic spline over 2 pixels on each side. We estimate the effective spatial resolution of the map to be somewhat larger than the resolution at 14.0\mic, i.e., $FWHM\gtrsim1.8\,{\rm px}=3.7\arcsec$. 

\begin{figure}[b!]
\includegraphics[angle=0,scale=0.8]{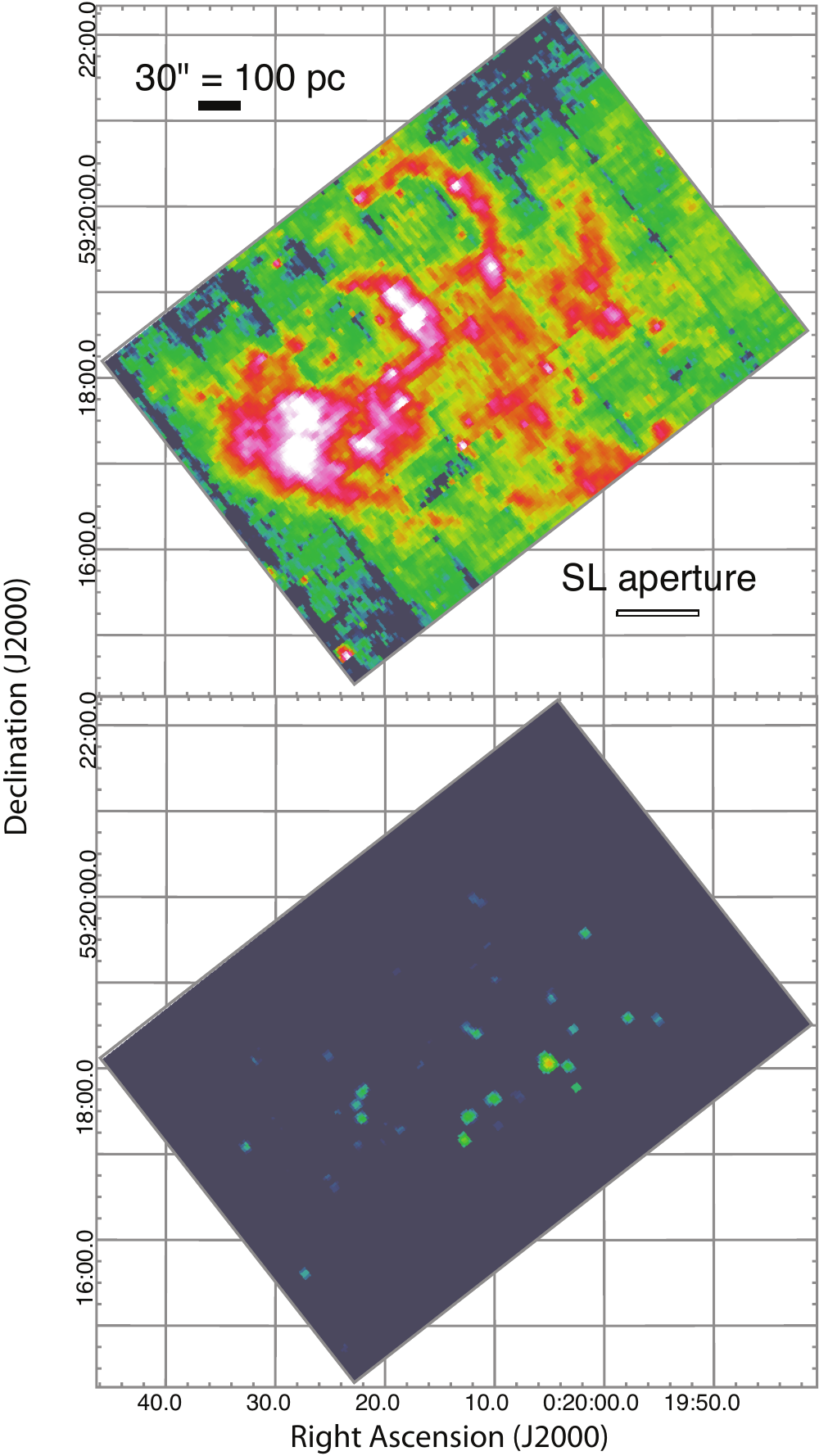}
\caption{Spitzer/IRS map. The top panel shows the distribution of the flux integrated along $7.4-14.5$\mic. The effective spatial resolution of the map ($\gtrsim3.2\arcsec$) is approximately equal to the width of the SL aperture. The silicate strength map is presented in the bottom panel (Section\,\ref{sec:map}). 
}
\label{fig:SL1map}
\end{figure}

A preliminary automatic cleaning of each exposure was performed using IRSCLEAN\footnote{Version 1.9; \textit{http://irsa.ipac.caltech.edu/data/SPITZER/docs/}}. The data were then imported and analyzed with CUBISM (version 1.7; Smith et al.\ 2007). A second manual cleaning step was performed using the backtracking tool provided by CUBISM. Several exposures at the edges of the map were chosen to remove the background emission, mostly arising from the Milky Way. Images corresponding to relevant wavelength ranges for building the silicate strength map (Section\,\ref{sec:map}) were extracted with CUBISM. The map of the integrated flux in the mid-IR range is shown in Figure\,\ref{fig:SL1map}.
%Based on the FWHM of sources in Figure\,\ref{fig:images}a, the spatial resolution is between $3\arcsec$ and $4\arcsec$, i.e., about 10-14\,pc at a distance of 700\,kpc.

Near-infrared photometry of stars toward IC\,10 is taken from the 2MASS point-like source catalog (Skrutskie et al.\ 2006) while optical photometry is taken from the Local Group Galaxies Survey (LGGS; Massey et al.\ 2007). In addition, we measured the mid-IR photometry using IRAC (Fazio et al.\ 2004) and MIPS (Rieke et al.\ 2004) onboard Spitzer. IC\,10 was observed with IRAC on 2004 July  23 with the 4 channels, centered respectively at $3.6$\mic, $4.5$\mic, $5.8$\mic, and $8.0$\mic\ (AORKEY 4424960), and with MIPS on 2004 December 26 at $24$\mic, $70$\mic, and $160$\mic\ (AORKEY 4425472). Point-like sources were identified with the software MOPEX\footnote{\textit{http://ssc.spitzer.caltech.edu/postbcd/mopex.html}} in all IRAC bands and in the MIPS 24\mic\ band. 

The longer wavelength bands of MIPS were not used because of the low spatial resolution. Similarly, we did not make use of the observation by the WISE telescope (Wright et al.\ 2010) at $3.4$\mic, $4.6$\mic, $12$\mic, and $22$\mic, because of the coarser spatial resolution (from $6.1\arcsec$ to $12\arcsec$) as compared to IRAC ($1.66-1.98\arcsec$) and MIPS 24\mic\ ($6\arcsec$). 

The low Galactic latitude of IC\,10 ($-3.34^\circ$) results in a high extinction, $E(B-V)=0.81$ (Massey \& Armandroff 1995; Massey et al.\ 2007), implying that observations are strongly reddened, even in the near-IR.  Using the value of $E(B-V) = 0.81$, and assuming the total-to-selective extinction ratio $R_v = 3.05,$ we de-reddened the $UBVR$ observations following the prescription of Table 3.21 in Binney \&  Merrifield (1998).  Magnitudes from the {\it I} band to the IRAC [8.0] band were corrected for extinction using the power law prescription of Martin \& Whittet (1990), summarized in Glass (1999).

\section{Pixel-based analysis}\label{sec:map}

In order to identify dust enshrouded O-rich stars in IC\,10, a silicate strength map was built from the IRS map (Section\,\ref{sec:obs}). Silicate dust is looked for in emission via the 9.7\mic\ emission feature which originates from the Si$-$O bond stretching mode (e.g., Knacke \& Thomson 1973). Following Spoon et al.\ (2007), the silicate strength $S_{\rm sil}$ is defined as:
\begin{equation}\label{eq:ssil}
S_{\rm sil} = \ln \frac{f_{\rm 9.7, obs}}{f_{\rm 9.7, cont}},
\end{equation}
where $f_{\rm 9.7, obs}$ is the observed flux density at $9.7$\mic, and $f_{\rm 9.7, cont}$ is the continuum flux density at the same wavelength. We first calculated the continuum shape in order to estimate $f_{\rm 9.7, cont}$. We used CUBISM to extract the continuum maps at $\sim5.4$\mic\ (median flux within $5.2$-$5.6$\mic) and $14.0$\mic\ (median flux within $13.7$-$14.3$\mic), and degraded the $5.4$\mic\ map to reach the spatial resolution of the $14.0$\mic\ map ($3.7\arcsec$). The continuum shape was then calculated by applying the spline method adapted to PAH-dominated spectra, as described by Spoon et al.\ (2007). Finally, we extracted the $9.7$\mic\ map and convolved it to a $3.7\arcsec$ resolution in order to estimate $f_{\rm 9.7, obs}$. We calculated $S_{\rm sil}$ using Equation\,\ref{eq:ssil} for all the pixels with a continuum flux at 9.7\mic\ greater than 1\,mJy.

\begin{figure*}
\begin{center}
\includegraphics[angle=0,scale=0.6]{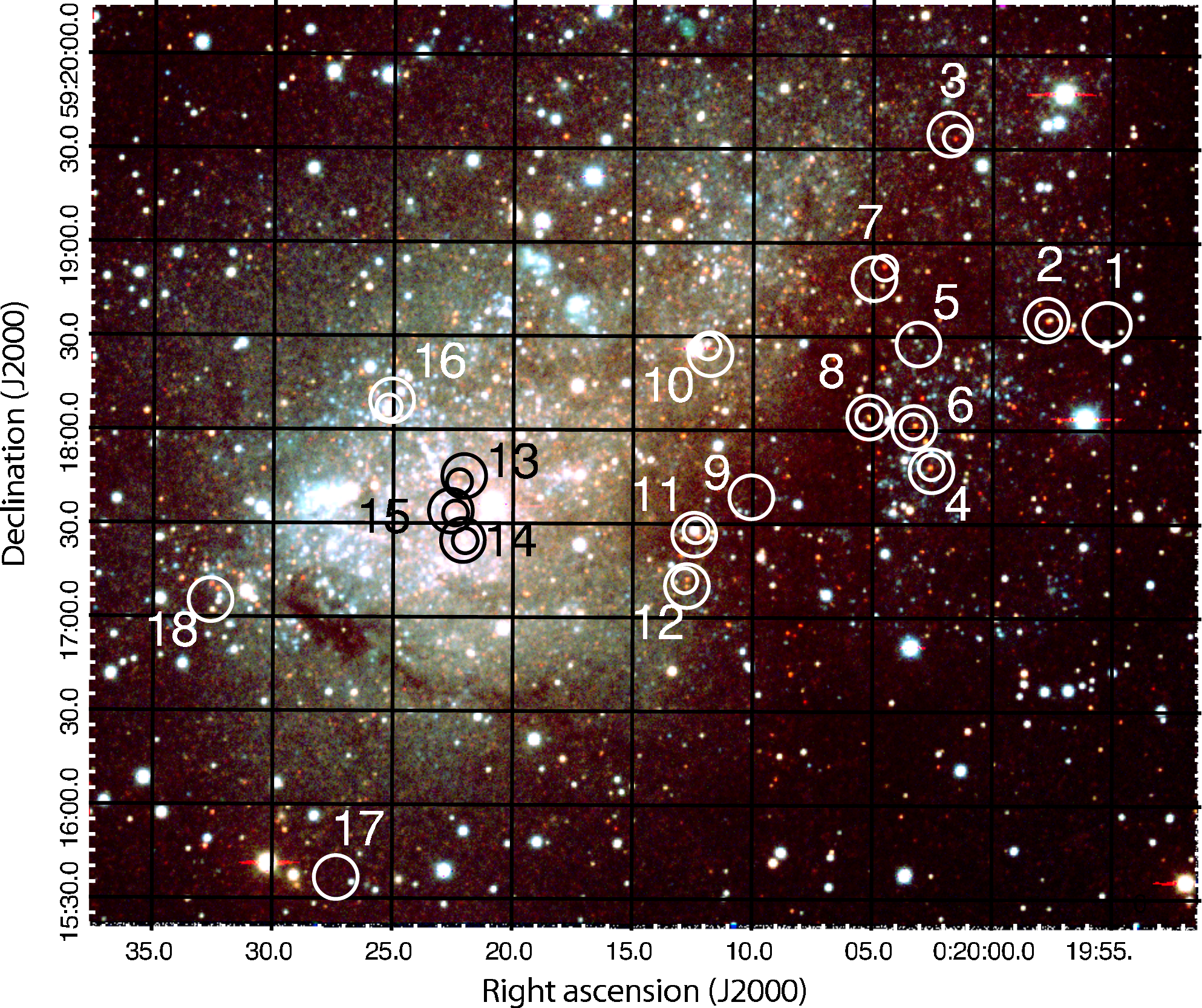}
\caption{Optical image of IC\,10, with I, B, and V band images as RGB colors. The outer circles ($7\arcsec$ radius; $\approx24$\,pc) are centered on candidate silicate emission point-like sources (Table\,\ref{tab:sources}). The inner circles ($4\arcsec$ radius; $\approx14$\,pc) indicate the position of the 2MASS counterpart (Table\,\ref{tab:photo_2mass}). The size of the circles is chosen for display purposes. The IRS spectra are extracted from a beam with FWHM$\approx4\arcsec$.  }
%\caption{Image of IC\,10 with RGB colors as Spitzer/24\mic\ (red), optical R band (green), and silicate strength (blue). The candidate silicate emission point-like sources are identified with thick white circles ($7\arcsec$ radius) and are sorted by ascending RA. The unlabeled thin dashed white circles indicate the distribution of WR stars (Massey et al.\ 1992; Massey \&\ Homes 2002; Royer et al.\ 2001).}
\label{fig:map}
\end{center}
\end{figure*}

The resulting silicate emission map is shown in Figure\,\ref{fig:SL1map}. Based on the FWHM of sources in Figure\,\ref{fig:map}, the spatial resolution ranges from $3\arcsec$ to $4\arcsec$, i.e., about $10-14$\,pc at a distance of 700\,kpc. A total of 18 point-like sources were selected based on their point-like appearance and for which $S_{\rm sil}>0.05$ (Table\,\ref{tab:sources}; Figure\,\ref{fig:map}). We find no evidence of spatial clustering and no evidence of extended silicate emission. 

We wish to emphasize that the silicate strength values inferred from the pixel-based analysis (Figure\,\ref{fig:SL1map}) are only indicative and bear systematic uncertainties. The map interpolation combined with the convolution to homogenize the data necessarily results in uncertain pixel values (Section\,\ref{sec:obs}). Most importantly, significant background emission arising from the ISM of IC\,10 prevents an accurate estimate of the source silicate dust continuum. Final values of the silicate strength are derived using optimal spectral extraction (Section\,\ref{sec:spectra}).

It is difficult to estimate the sample completeness based on the silicate strength map alone. Given the presence of gaps in the map (Section\,\ref{sec:obs}) and given that the point spread function is slightly more extended than the SL slit height ($\approx3.7\arcsec$), we could be missing somewhat less than half of the sources. The present sample is mostly limited by the flux at 10\mic\ (see also Section\,\ref{sec:similar}).

\begin{table}
\begin{center}
  \caption{Candidate silicate emission sources.}
  \label{tab:sources}
  \begin{tabular}{llll}
  \hline
ID &   $\alpha$ (J2000), $\delta$ (J2000) &  ID &    $\alpha$ (J2000), $\delta$ (J2000)  \\ %SIMBAD/optical image\\
  \hline
\#1      &  00:19:55.1, +59:18:34.7 &  \#10 &  00:20:11.8,  +59:18:25.5    \\
\#2 &  00:19:57.8,  +59:18:36.6 &  \#11 &  00:20:12.4, +59:17:26.3   \\%, HII region [HL90]7 at 11\arcsec / stars in optical image   \\
\#3 &  00:20:01.8, +59:19:34.9 & \#12 &  00:20:12.8,  +59:17:10.6        \\% / Carbon star [DBL2004]162 at 2.45\arcsec  / faint stars in optical image \\
\#4 &  00:20:02.6,  +59:17:46.6 &  \#13 &  00:20:22.1,  +59:17:43.9     \\% /  nothing within 15\arcsec / stars in optical image     \\
\#5 & 00:20:02.9, +59:18:27.3  & \#14 & 00:20:22.1, +59:17:24.9  \\
\#6 &  00:20:03.5,  +59:18:01.4 &  \#15 & 00:20:22.6, +59:17:34.6   \\%Carbon star [DBL2004]168 at 9.24\arcsec and WR star RSMV5 at 9.71\arcsec  / stars in optical image     \\
\#7 & 00:20:04.5, +59:18:50.0  &  \#16  &  00:20:25.1, +59:18:08.8   \\
\#8 &  00:20:04.9,  +59:18:03.3 &  \#17 & 00:20:27.3, +59:15:37.0        \\% / WR star RSMV5 at 6.6\arcsec 	 \\
\#9 &  00:20:10.1,  +59:17:39.3 & \#18 & 00:20:32.6, +59:17:05.7   \\%nothing within 19\arcsec  / faint star in optical image\\
  \hline
  \end{tabular}
%\tablefoot{$S_{\rm sil}$ is the value (in magnitudes) inferred from the silicate strength map. These values are simply informative, and should be regarded with caution (see text).}
%\tablenotetext{b}{Value (in magnitudes) based on the integrated CUBISM spectrum.}
%\tablenotetext{b}{The distance between the centroid in the map and the 2MASS coordinates is shown between parentheses.}
\end{center}
\end{table}

\section{Optical and infrared counterparts}\label{sec:sample}

%\subsection{Infrared and optical counterparts}\label{sec:counterparts}

%not 1, 5, 9, 17
\subsection{2MASS catalog}

Most candidate silicate emission sources in Table\,\ref{tab:sources} could be associated with a 2MASS source from the point-like source catalog (Table\,\ref{tab:photo_2mass}). Given the spatial resolution of the silicate strength map ($3\arcsec-4\arcsec$; Section\,\ref{sec:map}), we considered positive matches for association of $4\arcsec$ or less between the source centroid in the silicate strength map and the 2MASS coordinates. Most sources were matched within $2\arcsec$ or less, corresponding to about $6.8$\,pc in actual distance. There were no multiple matches within $4\arcsec$. We consider from now on the 2MASS coordinates as  our reference coordinates for the candidate silicate emission sources. 

Sources \#1, \#5, \#9, \#17, and \#18 could \textit{not} be associated with 2MASS point-like sources. Source \#18 is $4.4''$ away from 2MASS\,00203222+5917091 in the extended source catalog. Multiple sources from the LGGS (optical) catalog are seen close to this source, but no match in the IRAC [8.0] band was found (Section\,\ref{sec:irac}). For this reason, and because the coordinate match between source \#18 and the 2MASS counterpart is somewhat larger than the map resolution ($\approx4"$), we exclude this source from the following discussion. The match between the other sources and IRAC sources is discussed in Section\,\ref{sec:irac}.

\begin{table*}
\begin{center}
  \caption{2MASS Photometric data of associated stars.}
  \label{tab:photo_2mass}
  \begin{tabular}{l l c c c}
  \hline
ID &  2MASS ID                    & $J$ &$H$ & $K$                 \\ %SIMBAD/optical image\\
  \hline
\multicolumn{5}{c}{Foreground stars}\\
\hline
\#2 &   00195768+5918349 ($1.9\arcsec$) & ($15.19$) &  ($14.44\pm0.06$) & ($13.71$)   \\
\#10 & 	00201183+5918267  ($1.3\arcsec$)  & ($13.57$) & $15.39\pm0.16$ &  $14.75\pm0.16$  \\
\#11 &  00201237+5917279   ($1.7\arcsec$) & $13.62\pm0.03$ &  $12.56\pm0.04$ &  $12.08\pm0.03$ \\%nothing    \\
\#15 & 	00202240+5917332  ($2.0\arcsec$)  & $15.44\pm0.07$ &  $14.60\pm0.06$ &  $14.02\pm0.07$   \\ %J002022.52+591732.9 at 0.94\arcsec
\#16 & 	00202520+5918070	($1.9\arcsec$)    & $14.95\pm0.08$ &  $13.96\pm0.08$ &  ($13.18$)  \\
\hline
\multicolumn{5}{c}{IC\,10}\\
\hline
\#3 &  	00200155+5919332 ($2.5\arcsec$)  & $15.66\pm0.08$ & $14.60\pm0.06$ & $14.01\pm0.07$  \\%J002001.84+591933.9 at 2.1\arcsec, another at 2.9\arcsec \\
\#4 &  	00200259+5917481 ($1.5\arcsec$)&  $15.23\pm0.06$ & $14.15\pm0.05$ &  $13.56\pm0.05$ \\%J002002.61+591748.2 at 0.17\arcsec  \\
\#6 &  	00200322+5918013 ($2.1\arcsec$) &   $15.19\pm0.06$ &  $14.05\pm0.06$ &  $13.67\pm0.06$  \\%J002003.23+591801.6 at 0.30\arcsec     \\
\#7 &  00200452+5918521  ($3.6\arcsec$)  & $15.57\pm0.06$ &  $14.4\pm0.07$ &  $13.79\pm0.04$    \\ %J002004.54+591852.3 at 0.19\arcsec
\#8 &   00200510+5918039  ($1.7\arcsec$) & $15.60\pm0.08$ &  $14.81\pm0.10$ &  $14.13\pm0.07$ \\% J002005.11+591804.1 at 0.19\arcsec 	 \\
\#12 &  00201270+5917121  ($1.7\arcsec$) & $16.06\pm0.11$ &  $14.88\pm0.10$ &  $14.51\pm0.10$   \\%J002012.73+591712.3 at 0.20\arcsec      \\
\#13 &  00202225+5917432 ($1.4\arcsec$) & $15.59\pm0.09$ &  $14.47\pm0.09$ & $13.76\pm0.07$   \\%J002022.28+591743.3 at 0.19\arcsec   \\
\#14 & 	00202199+5917244 ($0.9\arcsec$) & $14.72\pm0.05$ &  $13.60\pm0.05$ &  $13.24\pm0.05$  \\ %J002022.01+591724.5 at 0.13\arcsec
  \hline
  \end{tabular}
\tablefoot{Sources \#1, \#5, \#9, \#17, and \#18 could not be matched with any 2MASS sources. Magnitudes between parentheses indicate upper limits or uncertain measurements. The field stars are identified  based on the color diagnostics discussed in Section\,\ref{sec:fore}. The distance between the source centroid in the silicate strength map and the associated 2MASS catalog is indicated between the parentheses. }
%\tablenotetext{a}{The distance between the source centroid in the map and the 2MASS coordinates is shown between parentheses.}
%\tablenotetext{b}{The distance between the 2MASS and LGGS coordinates is shown between parentheses.}
%\tablenotetext{c}{}
\end{center}
\end{table*}

\subsection{IRAC and MIPS sources}\label{sec:irac}

We cross-correlated our sample with sources identified in the IRAC maps with MOPEX (Section\,\ref{sec:obs}). At the spatial resolution of the IRAC [3.6] map ($1.66''$ or $\approx5.6$\,pc), sources \#4, \#8, \#11, and \#12 looked by eye slightly elongated while sources \#10 and \#16 are multiple or visibly extended. Based on the radii fitted by MOPEX, the other sources are point-like. IRAC magnitudes are presented in Table\,\ref{tab:irac}. 

\begin{figure*}
\begin{centering}
\includegraphics[angle=0,scale=0.9]{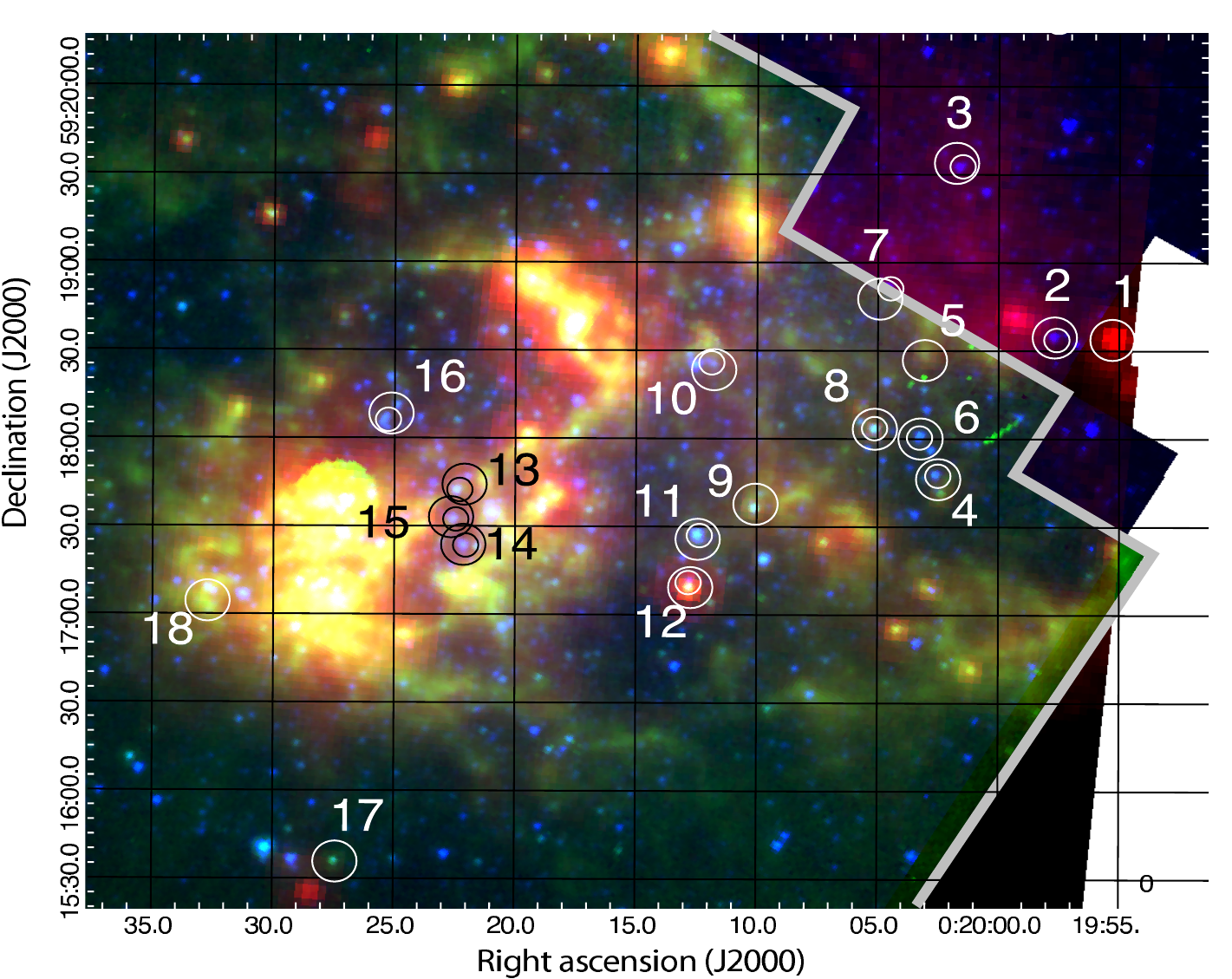}
\caption{Spitzer image (R: MIPS [24], G: IRAC [8.0], B: IRAC [3.6]). See Figure\,\ref{fig:map} for the symbol description.  
Sources \#8, \#11, and \#12 are 24\mic\ point-like sources. The gray polygon shows the area covered in all bands. 
}
%2MASS and Spitzer/IRAC images. The dotted rectangle is the Spitzer/IRS SL slit ($3.7\arcsec\times57\arcsec$, or $\approx12.6$\,pc$\times193$\,pc). The position of the extracted source is indicated by the segments on both side of the slit. The star indicates the center of the slit. }
\label{fig:spitzer}
\end{centering}
\end{figure*}

Source \#17 is not a 2MASS source, but it could be matched with an IRAC point-like source. It is likely significantly embedded so that even the near-IR bands are extinguished by dust. Thus, we choose to include this source in the following discussion. Sources \#1, \#5, and \#9 are either significantly far from any IRAC source or they are part of an extended emission in the IRAC bands. These $3$ sources are excluded in the following discussion since they also do not have a 2MASS point-like source counterpart.

%\begin{figure*}
%\includegraphics[angle=0,scale=0.7]{./figures/irac.pdf}
%\includegraphics[angle=0,scale=0.7]{./figures/mips.pdf}
%\caption{\textit{Top} $-$ IRAC [8.0\mic] image. \textit{Bottom} $-$ MIPS [24\mic] image.}
%\label{fig:images}
%\end{figure*}

\begin{table}
\begin{center}
  \caption{Spitzer photometry.}
  \label{tab:irac}
  \begin{tabular}{l l l l l l}
  \hline
ID &   [3.6] & [4.5] & [5.8] & [8.0] & [24] \\
  \hline
\multicolumn{6}{c}{Foreground stars}\\
\hline
 \#2 &  $13.31$ & ... & $13.01$ & ... & ...\\
\#10$^{\rm a}$ &  ... &  ... & $13.15$ & ... &...\\
\#11 &  $11.67$ & $11.68$ & $11.35$ & $10.97$ & $8.43\pm0.30$ \\
\#15 & $13.64$ & $13.55$ & $13.57$ & $13.15$ &...\\  
\#16$^{\rm a}$  & $12.94$ & $12.87$ & $12.55$ & ... &...\\  
\hline
\multicolumn{6}{c}{IC\,10}\\
\hline
 \#3 &  $13.32$ & ...$^{\rm b}$  & $12.59$ & ...$^{\rm b}$ &...\\
\#4 & $13.29$ & $13.16$ & $12.83$ & $12.47$ &...\\
\#6 & $13.05$ & $13.04$ & $12.82$ & $12.55$ &...\\
\#7 & $13.29$ & ...$^{\rm b}$ & $12.98$ & ...$^{\rm b}$ &... \\
\#8 &  $13.21$ & $12.65$ & $12.15$ & $11.13$ & $7.62\pm0.30$ \\
\#12 & $13.74$ & $12.96$ & $12.23$ & $10.77$ & $5.90\pm0.10$ \\
\#13 &  $13.20$ & ... & $12.40$ & ...&...\\  
\#14 & $13.12$ & $13.16$ & $12.37$ & ...&...\\  
\hline
\multicolumn{6}{c}{Unknown membership}\\
\hline
\#17 & $14.99$ & $13.81$ & $12.90$ & $12.07$ &...\\
  \hline
  \end{tabular}
\tablefoot{Magnitudes calculated assuming zero-magnitude fluxes from the instruments handbooks. The uncertainties are $\approx0.01$, $\approx0.01$, $\approx0.02$, and $\approx0.03$ for [3.6], [4.5], [5.8], and [8.0] respectively.
Bands with no data correspond to sources that were not found by the point source detection algorithm in MOPEX unless otherwise noted. The membership is based on the color diagnostics discussed in Section\,\ref{sec:fore}. }
\tablefoottext{a}{Blended or multiple objects.}
\tablefoottext{b}{Not covered by the IRAC observation.}
%\tablenotetext{b}{The distance between the centroid in the map and the 2MASS coordinates is shown between parentheses.}
\end{center}
\end{table}

Sources \#12 and, to a lesser extent, \#8 and \#11 are matched with MIPS 24\mic\ point-like sources (Figure\,\ref{fig:spitzer}). The spatial resolution at 24\mic\ is $6\arcsec$ (corresponding to $\approx20.4$\,pc), i.e., somewhat larger than the silicate strength map resolution ($\approx3.5\arcsec$, Section\,\ref{sec:map}). We verified that there were no multiple IRAC sources contributing to the extraction aperture at 24\mic.

\subsection{LGGS catalog}\label{sec:lggs}

Matches with the optical LGGS sample were found within $\sim1\arcsec$ of the 2MASS coordinates for all sources in our remaining sampled (Table\,\ref{tab:photo_opt}) except for sources \#3 and \#11. Because of the large IRS and IRAC beams, we consider that LGGS sources within $\sim4\arcsec$ could contribute to the extracted IR fluxes. We list in Tables\,\ref{tab:contam2} and \ref{tab:contam1} the LGGS sources found within $4''$ of the 2MASS counterpart. In practice, only sources \#12, \#13, \#14, and \#16 in our sample have several bright infrared LGGS stars within such a radius. It must be kept in mind in the following that these sources could be multiple objects.

\begin{table*}
\begin{center}
  \caption{Optical photometric data of associated stars.}
  \label{tab:photo_opt}
  \begin{tabular}{l l c c c c c}
  \hline
ID &   LGGS ID                     & $U$&  $B$ & $V$ & $R$ & $I$ \\ %SIMBAD/optical image\\
  \hline
\multicolumn{7}{c}{Foreground stars}\\
\hline
\#2 &    J001957.61+591835.5  ($0.8\arcsec$) &  ... &  $22.58$ &  $20.94$ &  $19.90$ &  $18.77$  \\%J001957.61+591835.5 at 0.8\arcsec  \\
\#10 & 	 J002011.91+591827.6  ($1.0\arcsec$)& $21.70\pm0.01$ &  $21.29$ &  $19.91$ &  $19.03$ &  $18.07$ \\%J002011.91+591827.6 at 1.02\arcsec    \\
\#11 &  ... & ... & ... & ... & ... & ... \\%nothing    \\
\#15 & 	    J002022.52+591732.9   ($0.9\arcsec$) &  $19.53$ &  $19.34$ &  $18.40$ &  $17.81$ &  $17.17$                            \\ %J002022.52+591732.9 at 0.94\arcsec
\#16 & 	  J002025.23+591807.3 ($0.5\arcsec$)  &  $19.89\pm0.02$ &  $19.96$ &  $20.86\pm0.06$ &  $19.45$ &  $18.50\pm0.01$ \\%J002025.23+591807.3 at 0.27\arcsec, another at 0.53\arcsec    \\
\hline
\multicolumn{7}{c}{IC\,10}\\
\hline
\#3 &  	 ... & ...& ... & ... & ... & ... \\%J002001.84+591933.9 at 2.1\arcsec, another at 2.9\arcsec \\
\#4 &   J002002.61+591748.2  ($0.2\arcsec$)& ... &  $24.67\pm0.02$ &  $21.63$ &  $19.62$ &  $17.43$   \\%J002002.61+591748.2 at 0.17\arcsec  \\
\#6 &  	 J002003.23+591801.6  ($0.3\arcsec$)& ... &  $24.02$ &  $21.27$ &  $19.43$ &  $17.39$   \\%J002003.23+591801.6 at 0.30\arcsec     \\
\#7 &   J002004.54+591852.3  ($0.2\arcsec$)& ... &  $25.29\pm0.08$ &  $22.56\pm0.02$ &  $20.35$ &  $17.92$    \\ %J002004.54+591852.3 at 0.19\arcsec
\#8 &    J002005.11+591804.1   ($0.2\arcsec$) & ... &  $23.35\pm0.02$ &  $21.10$ &  $19.43$ &  $17.71$ \\% J002005.11+591804.1 at 0.19\arcsec 	 \\
\#12 &   J002012.73+591712.3  ($0.2\arcsec$) & ... &  $24.54\pm0.02$ &  $22.02\pm0.01$ &  $20.30$ &  $18.32$   \\%J002012.73+591712.3 at 0.20\arcsec      \\
\#13 &   J002022.28+591743.3  ($0.2\arcsec$)& ... &  $23.92\pm0.01$ &  $21.68\pm0.01$ & $19.81$ &  $17.71$  \\%J002022.28+591743.3 at 0.19\arcsec   \\
\#14 &   J002022.01+591724.5  ($0.1\arcsec$)&  ... &  $22.74$ &  $20.10$ &  $18.52$ &  $16.89$  \\ %J002022.01+591724.5 at 0.13\arcsec
  \hline
  \end{tabular}
\tablefoot{Errors are below $0.01$\,dex unless otherwise noted. The field stars are identified  based on the color diagnostics discussed in Section\,\ref{sec:fore}. The distance between the source centroid in the silicate strength map and the associated LGGS catalog is indicated between the parentheses. The LGGS ID is the best match within the search radius (see Tables\,\ref{tab:contam2} and \ref{tab:contam1}).}
%\tablenotetext{a}{The distance between the source centroid in the map and the 2MASS coordinates is shown between parentheses.}
%\tablenotetext{b}{The distance between the 2MASS and LGGS coordinates is shown between parentheses.}
%\tablenotetext{c}{}
\end{center}
\end{table*}

\begin{figure*}
\begin{center}
\includegraphics[angle=0,scale=0.72]{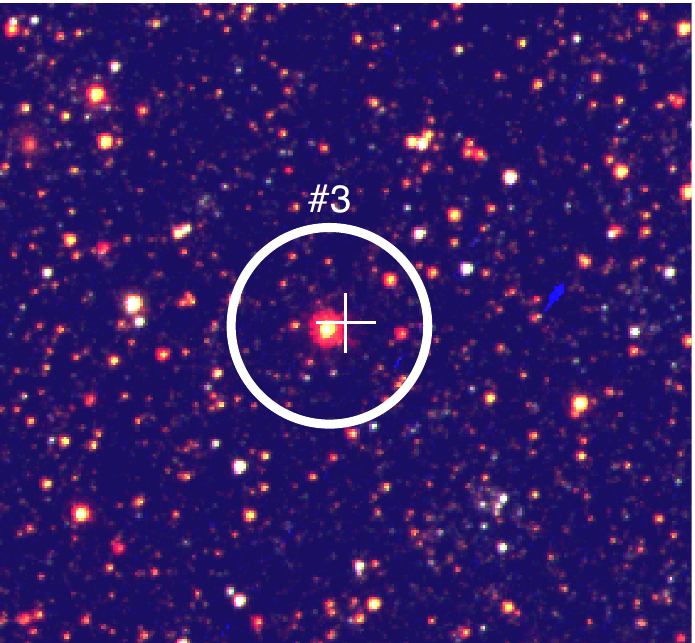}
\includegraphics[angle=0,scale=0.72]{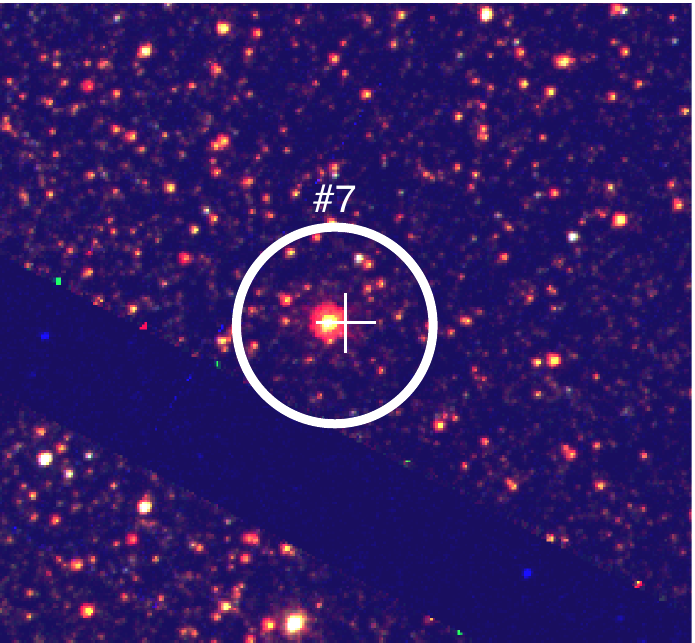}
\includegraphics[angle=0,scale=0.72]{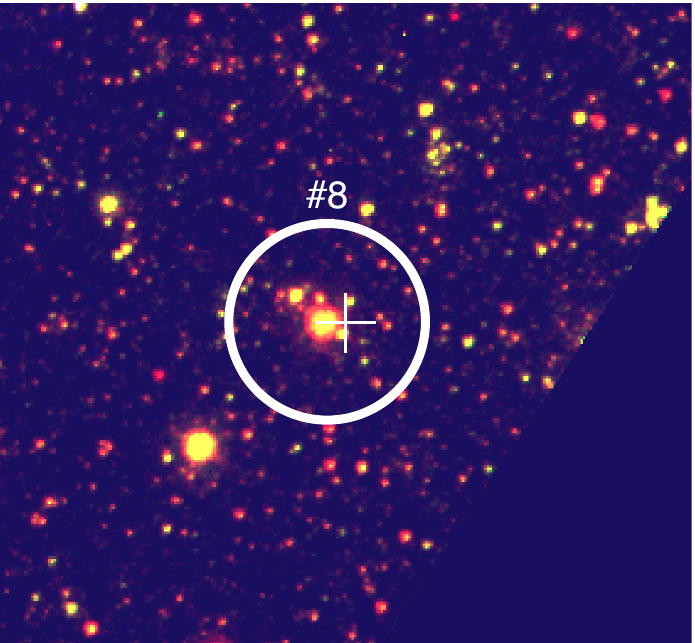}\\
\includegraphics[angle=0,scale=0.72]{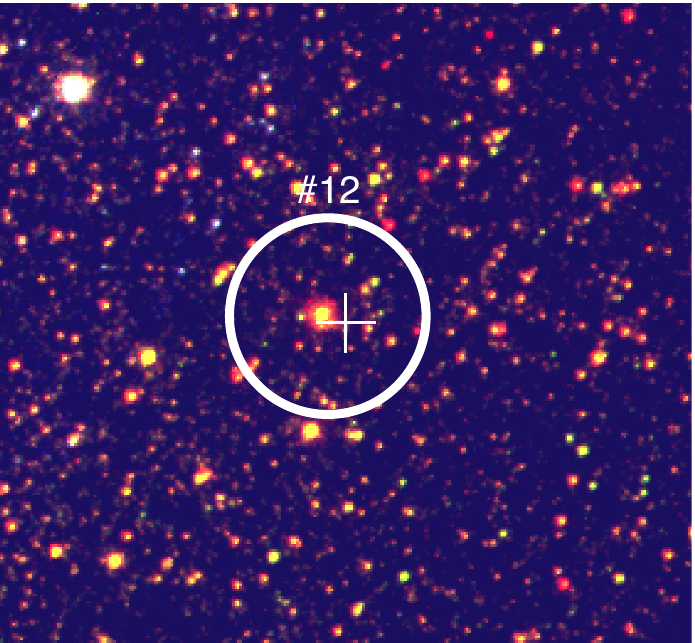}
\includegraphics[angle=0,scale=0.72]{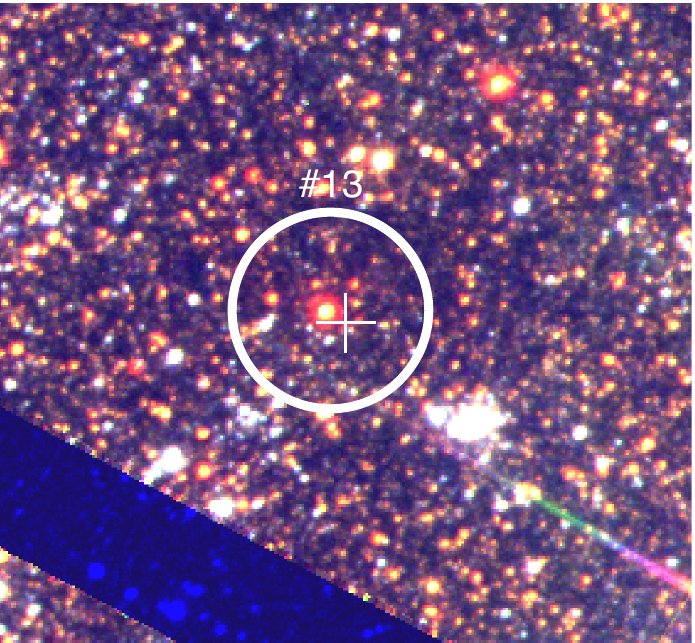}
\includegraphics[angle=0,scale=0.72]{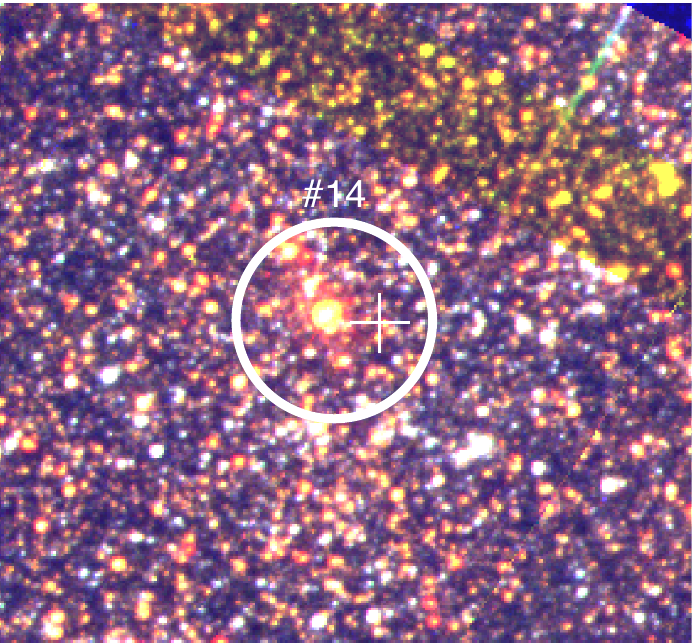}
\caption{HST/ACS images from candidate silicate emission sources within IC\,10. Images were downloaded from the Hubble Legacy Archive (\textit{http://hla.stsci.edu/}) with I, G, and B band images as RGB colors. The cross indicates the 2MASS coordinates. The circle represents the size of the IRS extraction aperture, $2\arcsec$ radius (corresponding to $\approx7$\,pc). The other sources were not observed with ACS. 
\label{fig:hst}}
\end{center}
\end{figure*}

For all the other sources, we cannot exclude that compact stellar clusters might be affected by confusion, even in the optical LGGS observations. Such clusters would have to be smaller than $\sim1\arcsec$, or $\sim3$\,pc, to be unresolved in the LGGS. We explored the high spatial resolution observations from the Hubble Space Telescope (HST) to investigate further the possible presence of compact stellar clusters. For this test, only sources that are member of IC\,10 are considered (Section\,\ref{sec:fore}). Only sources \#3, \#7, \#8, \#12, \#13, and \#14 were covered by the observations with the Advanced Camera for Surveys (ACS) (Figure\,\ref{fig:hst}). There is no evidence of enhanced clustering toward these sources, except maybe for source \#8, with a few bright stars within the IRS extraction aperture. In the following, we assume that the flux extracted in 2MASS, IRAC, and IRS is dominated by the LGGS object found closest to the 2MASS coordinates.

%2
%10 no clustering
%11 evidence of clustering
%15 no clustering, bright yellow star in beam
%16 evidence of clustering

%3 no clustering
%4
%6
%7 no clustering
%*8 slight clustering
%*12 no clustering
%*13 single red star dominates, cluster ~2" away
%*14 no clustering

%2: 3876: 13x17
%3: no IRAC
%4: 3729: 13x23
%6: 3683: 17x15
%7: 3200: 11x15
%8: 3519: 15x25
%10: too extended, not found as a PLS
%11: 3140: 17x23
%12: 3259/3260: 23x17, 13x9, could I have chosen the wrong one in IRAC
%13: 2113: 17x15
%14: 2297: 9x18
%15: 2211: 11x11
%16: 1629: 29x27
%17: 2669: 11x15

\subsection{Field contamination by foreground stars}\label{sec:fore}

\begin{figure}
\includegraphics[angle=0,scale=0.55]{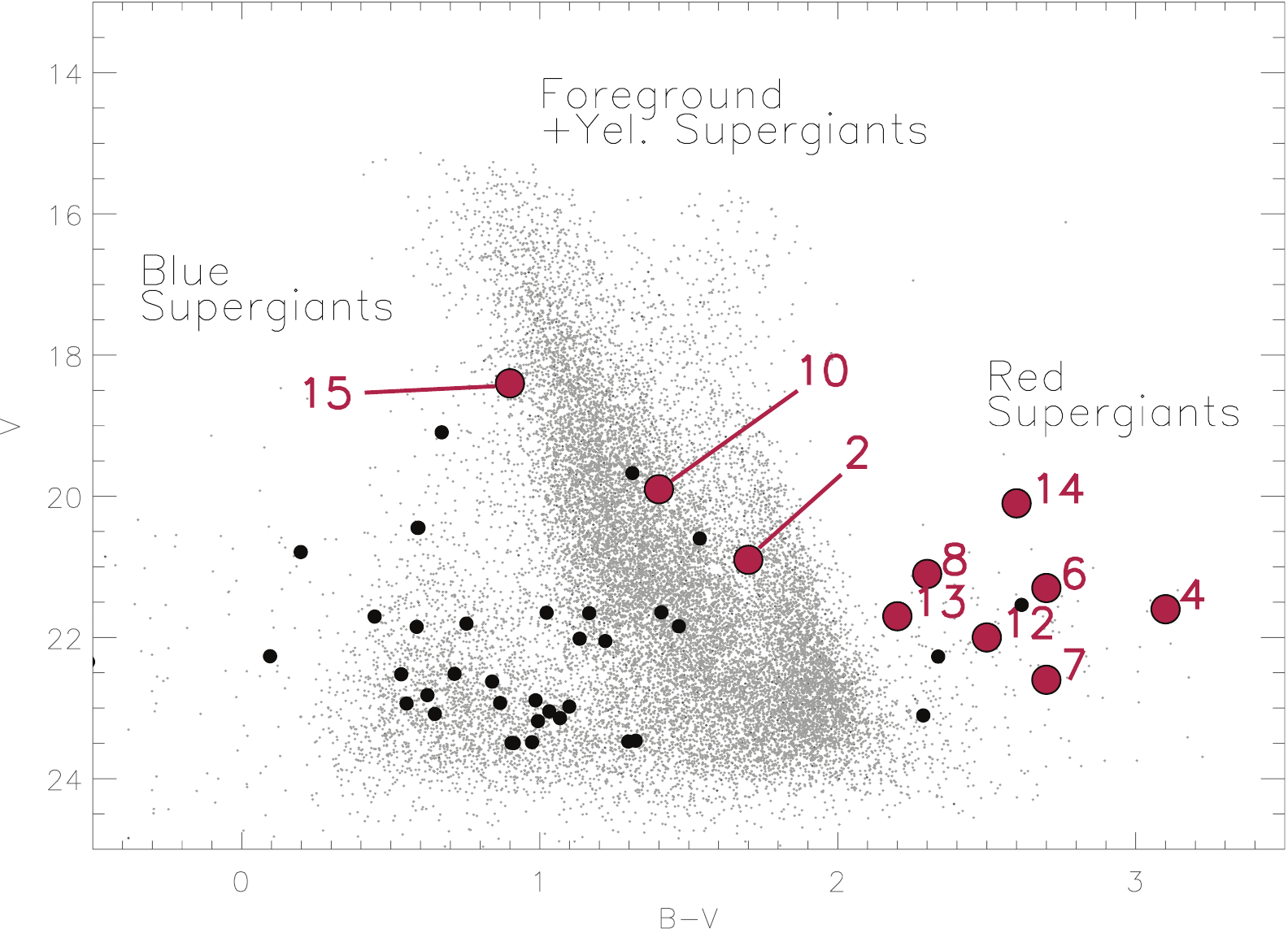}
\caption{$V$ vs.\ $B-V$ colors of stars toward IC\,10 (Massey et al.\ 2007). The candidate silicate emission sources from this study are shown as red points. Small black points indicate LGGS sources within the search radius of the candidate silicate emission sources (Table\,\ref{tab:contam2}).
\label{fig:cola}}
\end{figure}

We now investigate the photometry of the objects associated with the candidate silicate emission sources (Tables\,\ref{tab:photo_2mass} and \ref{tab:photo_opt}) in order to test their membership to IC\,10 and their intrinsic stellar properties. As explained in Massey et al.\ (2007), the $V$ and $B-V$ colors provide a good diagnostic on the stellar type while also separating foreground stars from stars in IC\,10. Figure\,\ref{fig:cola} shows the photometric data from Massey et al.\ with the candidate silicate emission sources from Table\,\ref{tab:sources} overlaid. According to Massey et al., RSGs belonging to IC\,10 are expected to have $B-V\gtrsim2$ and $V\lesssim20$. Only sources \#4, \#6, \#7, \#8, \#12, \#13, and \#14 fit this criterion. The other sources (\#2, \#10, and \#15) could be yellow supergiants, but they are far more likely foreground stars. Note that sources \#3 and \#11 have no optical counterparts (Section\,\ref{sec:lggs}). Diagnostics for these 2 sources are based on their IR photometry alone. We also show in Figure\,\ref{fig:cola} the LGGS sources found within the search radius of the candidate silicate emission sources from Table\,\ref{tab:contam2}. Only 3 sources have optical colors expected from RSGs, but they are not the most infrared bright within the search radius.  

In an independent approach, we have used the Besan\c{c}on population synthesis model (Robin et al.\ 2003) to count the number of foreground stars expected toward IC\,10 with colors and magnitudes similar to those expected from RSGs in IC\,10. We used the expected $V$ magnitude and $B-V$ color of RSGs in IC\,10 from Massey et al.\ (2007) and found that no foreground stars can be at the same time that red and that bright.

%The diagram $B-V$ vs.\ $V-R$ enables another diagnostic. Two branches are observed with the foreground stars occupying the lower branch and RSGs the upper (see the application in M\,31 by Massey et al.\ 2009). Effect of metallicity and reddening are discussed in Massey (1998) and Massey et al.\ (2009).
%We show the photometric data points of IC\,10 in Figure\,\ref{fig:colb}. The split between the 2 branches is barely seen, starting around $V-R\sim1.3$. Sources \#2, \#5, and \#10 are located with foreground stars. Sources \#8 and \#13 are located between the 2 branches, while the other sources are located in the RSG branch. 

%\begin{figure*}
%\epsscale{1.}
%\plotone{./figures/bv_vr.ps}
%\caption{$B-V$ vs.\ $V-R$ colors of stars toward IC\,10. Only stars with $V<21.5$ are shown. Red points indicate the candidate silicate emission sources from this study.} \label{fig:colb}
%\end{figure*}

Finally, near-infrared photometry further constrains the membership to IC\,10. Based on an offset field of view, Borissova et al.\ (2000) conclude that foreground stars have colors such that $0.4\lesssim J-K \lesssim1.0$, and $H-K \lesssim0.1$. All the sources in Table\,\ref{tab:photo_2mass} have $H-K>0.1$ and $J-K$ between $1.4$ and $1.8$, which bolsters our confidence that sources \#4, \#6, \#7, \#8, \#12, \#13, and \#14 belong to IC\,10. Moreover, Borissova et al.\ estimate that RSGs should have $J-K\sim1.4$ and $13\lesssim K \lesssim15$ while AGB stars should be fainter. Source \#11, which has no optical counterpart, has $K=12.1$ and does not fit the constraints above; it is brighter in the $K$ band by one order of magnitude than the other sources and is likely a foreground star. All the other sources have $1.4<J-K<1.8$. In particular, source \#3, which also has no optical counterpart, could be a RSG in IC\,10 based solely on its $JHK$ colors. 

In summary, both the optical and IR photometry of sources \#4, \#6, \#7, \#8, \#12, \#13, and \#14 are consistent with membership in IC\,10. Furthermore, the $K$ magnitude of these sources (and of source \#3) seem to indicate they are RSGs. We refine the determination of the stellar nature in Section\,\ref{sec:properties}.

\section{Mid-infrared spectra}\label{sec:spectra}

In this section, we present the mid-IR Spitzer/IRS spectra of all the candidate silicate emission sources. We consider sources \#3, \#4, \#6, \#7, \#8, \#12, \#13, and \#14 (members of IC\,10) as well as \#17 (IRAC source). 

\begin{figure}
\begin{center}
\includegraphics[angle=0,scale=0.43]{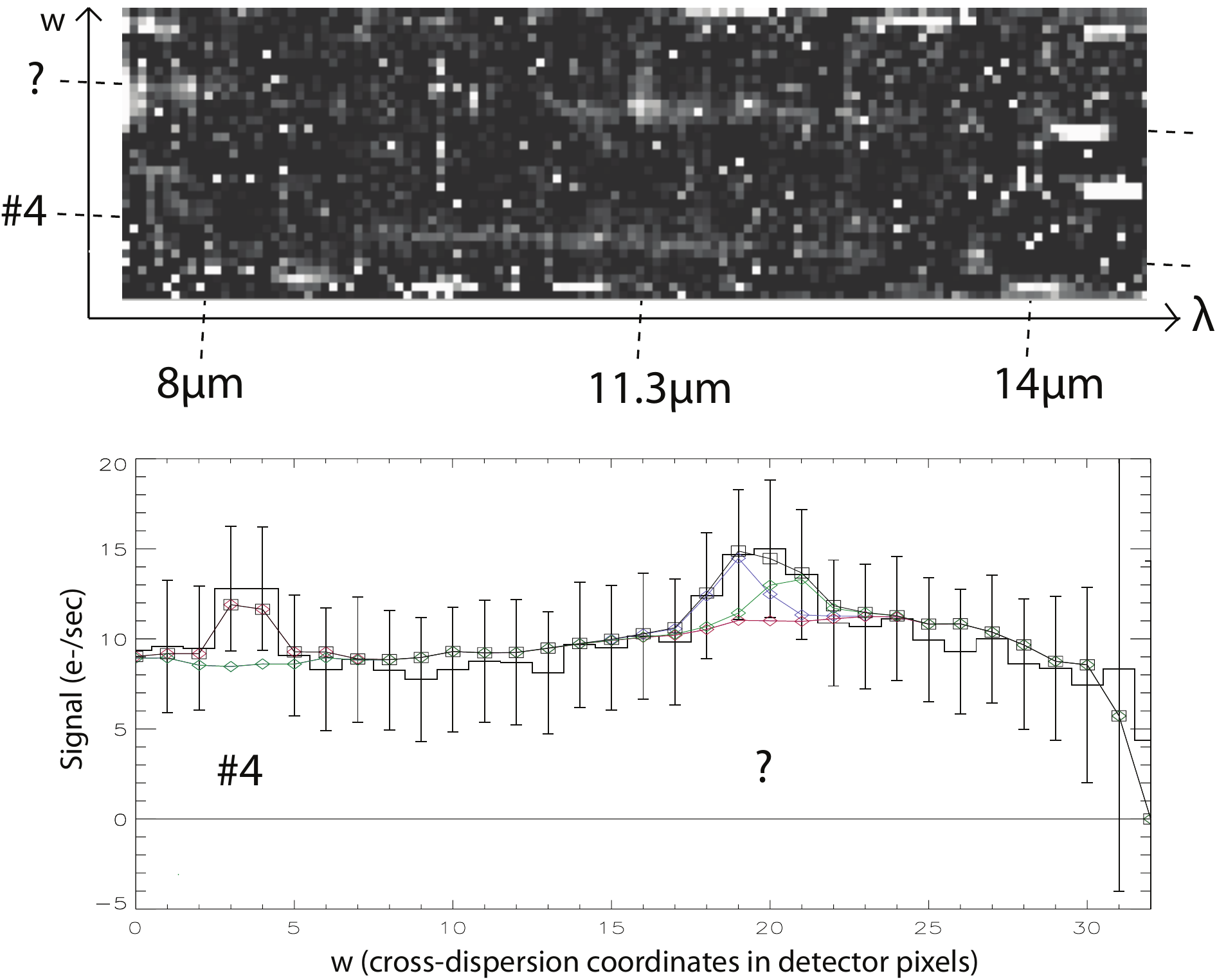}
  \caption{Example of optimal spectral extraction (source \#4). The detector image is shown in the top panel, \textit{after background subtraction}, with the cross-dispersion profile ($w$) as a function of wavelength. The corresponding profile along the aperture (integrated signal over the wavelength range vs.\ $w$) is shown as a histogram in the bottom panel. The connected squares show the fit of the spatial components in the slit, including source \#4 (red profile), the extended background emission, and another slightly extended source in IC\,10 matching the location of the H\2\ region [HL90]\,17 (Hodge \& Lee 1990), here fitted with 2 point-like sources showed by the green and blue profiles).  }
  \label{fig:extraction4}
  \end{center}
\end{figure}

\begin{figure*}
\includegraphics[angle=0,scale=0.99]{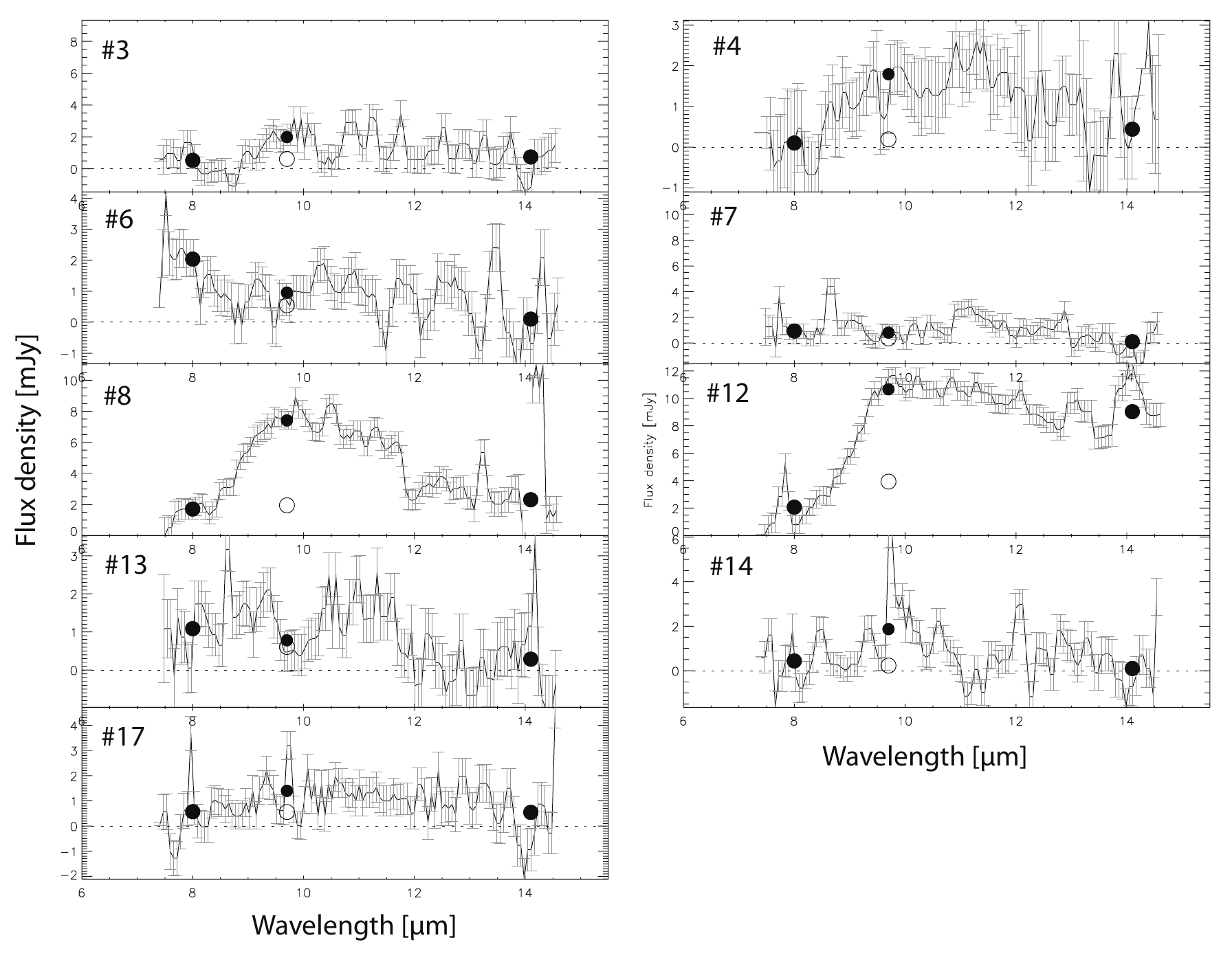}
  \caption{Spitzer/IRS spectra of sources photometrically identified as IC\,10 sources. The 3 filled circles represent the observed continuum flux density at 8.0\mic, 9.7\mic, and 14.0\mic\ while the open circle represents the interpolated flux density at 9.7\mic\ used to infer the silicate strength (see text). Spectra were smoothed by a running $3$-pixel median box.}
  \label{fig:opti}
\end{figure*}

The presence of spatially extended MIR emission (dominated by polycyclic aromatic hydrocarbon bands and warm dust continuum) from the ISM of IC\,10 prevents a regular spectral extraction of the sources (i.e., integration of the flux within a spatial window). We therefore used the optimal extraction provided by SMART-AdOpt\footnote{Version 8.2.4; \textit{http://isc.astro.cornell.edu/IRS/SmartRelease}} (Lebouteiller et al.\ 2010) to extract the spectra at the matching star location in the exposure images. Optimal extraction weighs the source spatial profile by using the instrument point spread function as a reference. The extended emission coming from the ISM of IC\,10 was removed simultaneously using a second- or third-order polynomial. The stars location within the IRS SL aperture was constrained not only in the cross-dispersion direction but also in the dispersion direction, therefore accounting for the slit throughput and providing an accurate flux calibration. An example of extraction is presented in Figure\,\ref{fig:extraction4}. Final spectra are presented in Figure\,\ref{fig:opti}. 

The spectral trace was detected for all the sources mentioned above. Detection levels (based on the integrated SL wavelength range) are given in Table\,\ref{tab:ssil}. Besides the bright sources \#8 and \#12, we note that sources \#4, \#6, \#13, and \#17 are fairly well detected (more than $2\sigma$), while sources \#3, \#7, and \#14 barely stand above the detection threshold. Based on the comparison between the source spatial profile and the IRS point spread function, we find that all sources are point-like at the spatial resolution of the IRS SL module at 10\mic, i.e., $\approx2''$. 

%Wm-2um-1 = 3e8/(0.54e-6)^2*1e-32 * S(Jy)
%Jy = Wm-2um-1 / (3e8/(0.54e-6)^2*1e-32) => 0.03 Jy -> 3Jy MW

Sources \#8 and \#12 show prominent silicate emission peaking respectively at $8$\,mJy, and $11$\,mJy, while source \#4 shows weak emission peaking at $\approx3$\,mJy (Figure\,\ref{fig:opti}). The signal-to-noise ratio of the other sources is too low to assert unambiguously the presence of silicate dust. Silicate strength values are given in Table\,\ref{tab:ssil}. 
The silicate emission flux density peak in RSGs of the Large Magellanic Cloud (LMC) ranges from $\sim1$\,Jy to $\sim3.5$\,Jy (Buchanan et al.\ 2009). Such sources would have fluxes around $4-14$\,mJy at the distance of IC\,10, which is compatible with our values. 
We notice that source \#12 is characterized by a bright dust continuum longward of $13$\mic\ which cannot be due to background extended emission, as it was removed during spectral extraction. This object is associated with a point-like source in the MIPS 24\mic\ image. The nature of the silicate emission sources is discussed in Section\,\ref{sec:model}.

%We conclude that at least sources \#4, \#8, and \#12 are associated with circumstellar silicate dust. 

%11.27 mag at 24um

%What are the photometric properties of sources showing silicate emission sources? Interestingly, \#8 and \#12 have similar colors: B-V~2.4, V-R~1.7, J-K~1.5, K~14.3.
%what about IRAC?

\begin{table}
\begin{center}
  \caption{Silicate strength values and model results.}
  \label{tab:ssil}
  \begin{tabular}{lcllll}
  \hline
ID & Detection\tablefootmark{a} &  $S_{\rm sil}$ &   $L_{\rm bol}$  & $T_{\rm eff}$ & MLR\tablefootmark{b} \\
  &        ($\sigma$)      & (mags) &   (L$_\odot$) & (K) & (M$_\odot$\,yr$^{-1}$) \\
  \hline
  \#3 & 1.4 & $+1.42^{+0.63}_{-1.47}$  & $120\,000$ & $3\,550$ & $8\times10^{-7}$   \\
  \textbf{\#4}  &3.4 & $+2.22^{+1.63}_{-2.15}$  & $130\,000$ & $3\,397$ & $2\times10^{-7}$ \\
   \#6 & 2.5 &$-0.53^{+0.75}_{-1.64}$ & $130\,000$ & $3\,490$ & $2\times10^{-7}$ \\
   \#7 & 1.5 & $-0.89^{+0.70}_{-2.34}$& $110\,000$ & $3\,397$ & $3\times10^{-7}$ \\
   \textbf{\#8} & 7.7 & $+1.44^{+0.07}_{-0.07}$& $120\,000$ & $3\,550$ & $28\times10^{-7}$ \\
   \textbf{\#12} & 18 &$+0.98^{+0.09}_{-0.10}$  & $90\,000$ & $3\,550$ & $30\times10^{-7}$ \\
   \#13 & 2.4 & $+0.85^{+0.65}_{-2.00}$  & $110\,000$ & $3\,550$ & $14\times10^{-7}$ \\
   \#14 & 1.3 & $+2.35^{+1.37}_{-2.49}$ & $150\,000$ & $3\,550$ & $5\times10^{-7}$ \\
   \#17 & 2.3 & $+1.27^{+0.85}_{-3.54}$ & $15\,000$ & $3\,550$ & $30\times10^{-6}$ \\
 \hline
  \end{tabular}
  \tablefoot{The sources with significant silicate emission are shown in bold (Section\,\ref{sec:spectra}). }
  \tablefoottext{a}{Detection level over the spectral trace (integrated SL wavelength range).}
  \tablefoottext{b}{Mass-loss rate determinations with a factor of $\approx2$ statistical uncertainties. }
%\tablecomments{Magnitudes calculated assuming zero-magnitude fluxes.}
%\tablenotetext{a}{Source out the field of view.}
%\tablenotetext{b}{Source not found by SExtractor.}
%\tablenotetext{b}{The distance between the centroid in the map and the 2MASS coordinates is shown between parentheses.}
\end{center}
\end{table}

\section{Properties of the stars}\label{sec:properties}

\subsection{RSG vs.\ AGB}\label{sec:model}

%  \caption{Model results. For each source, the SED model is shown on top as the solid curve, with the diamonds indicating photometry points from the optical bands and from 2MASS, and with the segments indicating the data. 
%  A close-up on the SL range is shown on bottom with the model fit as the dashed curve.}

%\subsection{Stellar nature}\label{sec:rsg_vs_agb}

\begin{figure}
\includegraphics[angle=0,scale=0.53]{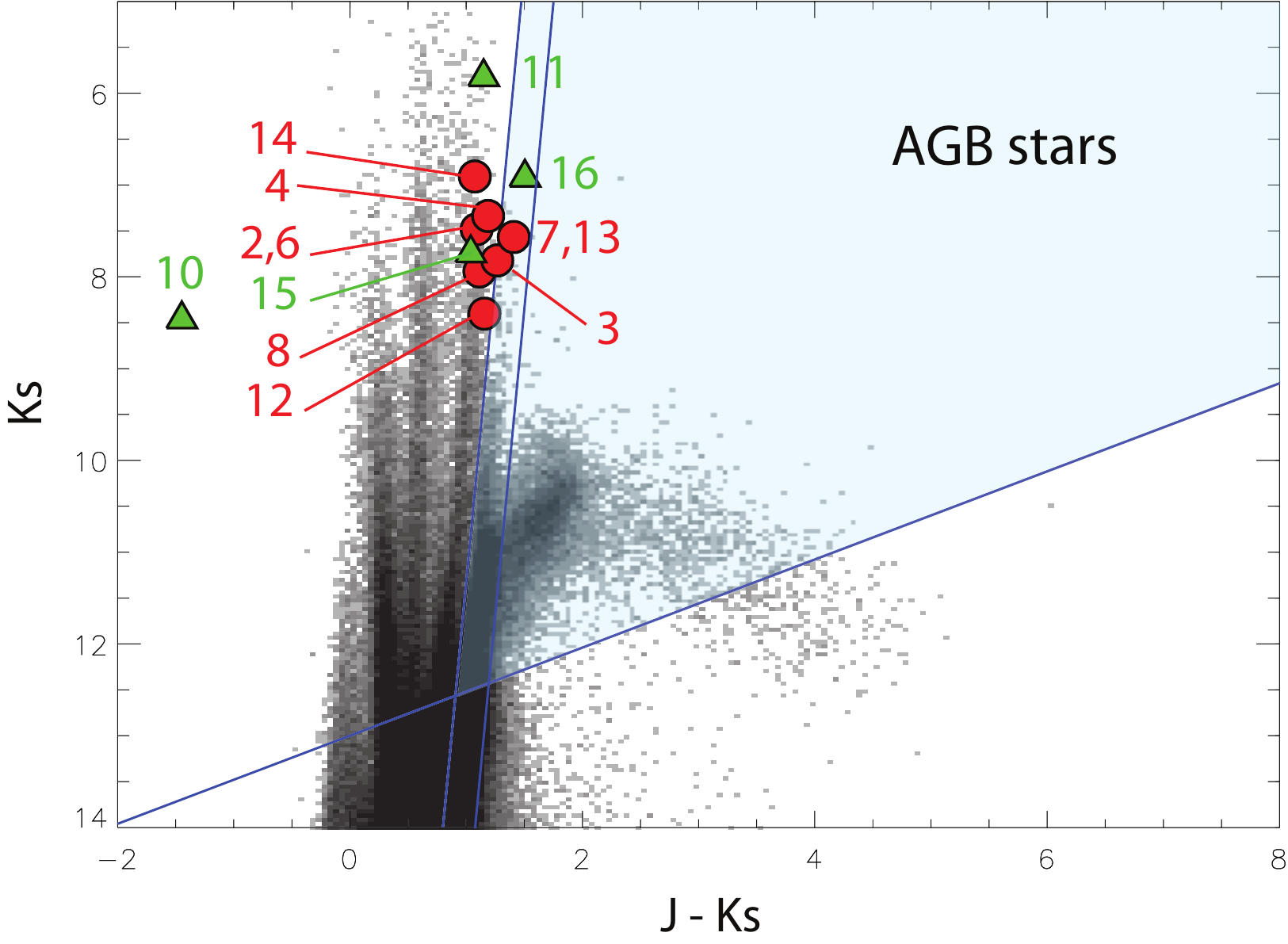}
\caption{$K_{\rm s}$ vs.\ $J-K_{\rm s}$ CMD for sources in IC\,10. The background is a Hess diagram of the sources from the SAGE catalog of the LMC (Meixner et al.\ 2006).  The IC\,10 sources are corrected for reddening and then "moved" to the LMC for comparison. AGB stars fall within the labeled region in the upper-right region (Cioni et al.\ 2006; Nikolaev \&\ Weinberg 2000). O-rich AGB stars are redder than the left oblique line while C-rich AGB stars are redder than the right oblique line. Sources \#10, \#11, \#15, and \#16 are foreground sources not associated with IC10, and are marked with green triangles. Source \#17 is not a 2MASS source.}
  \label{fig:ch34_jk}
\end{figure}

Silicate dust can be produced by several types of stars, most notably AGBs, RSGs, planetary nebulae, and novae. Young stellar objects (YSOs) also show silicate dust in their disks or outflows, although the dust grains might not be produced in situ. The optical and near-IR colors suggest that all sources but \#17 could be RSGs (Section\,\ref{sec:fore}). We now review this finding by comparing the $K_{\rm s}$ vs.\ $J-K_{\rm s}$ colors of stars in IC\,10 and in the LMC. Figure\,\ref{fig:ch34_jk} shows the CMD in which the magnitudes of IC\,10 sources have been scaled to the distance of the LMC. Sources \#7 and \#13 appear to fall in the O-rich AGB color domain, while source \#3 lies on the AGB/RSG cut. All the other sources (\#2, \#4, \#6, \#8, \#12, and \#14) are unlikely to be AGB stars. 

\begin{figure*}
\begin{center}
\includegraphics[angle=0,scale=0.65,clip=true]{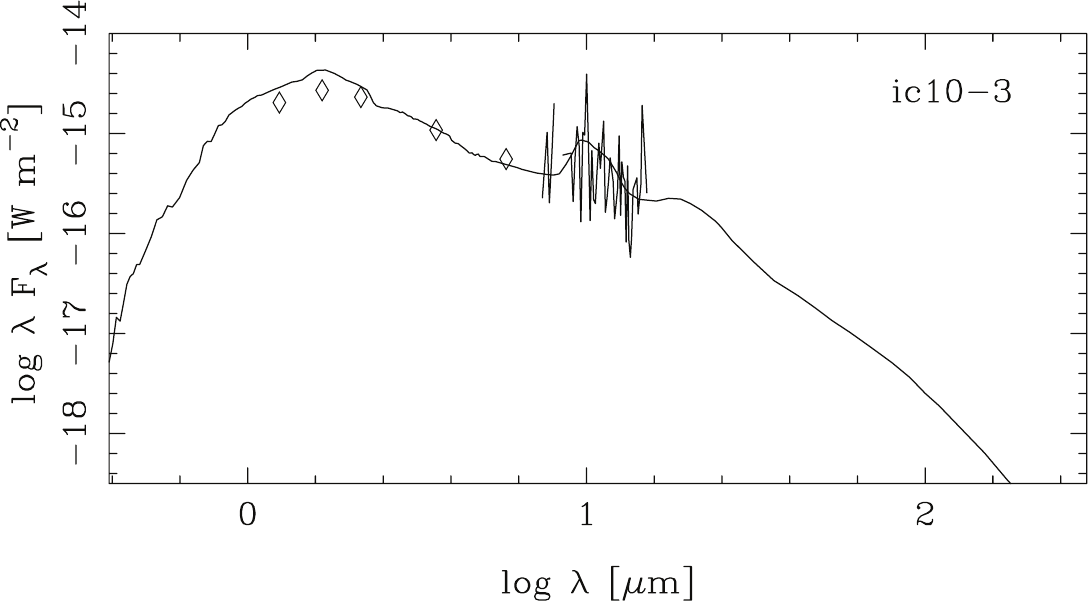}
\includegraphics[angle=0,scale=0.65,clip=true]{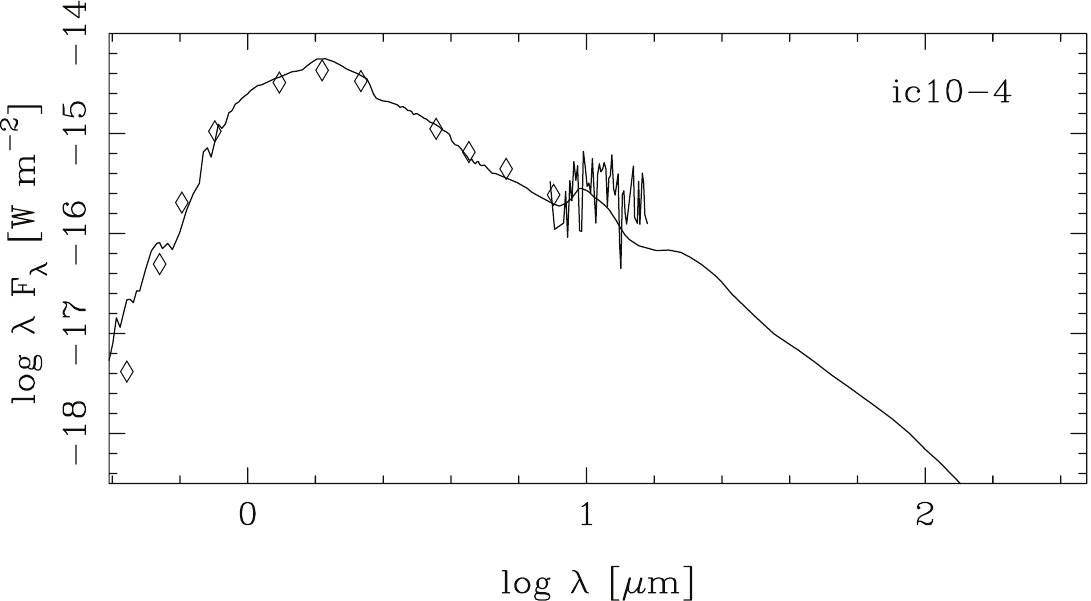}\\
\includegraphics[angle=0,scale=0.65,clip=true]{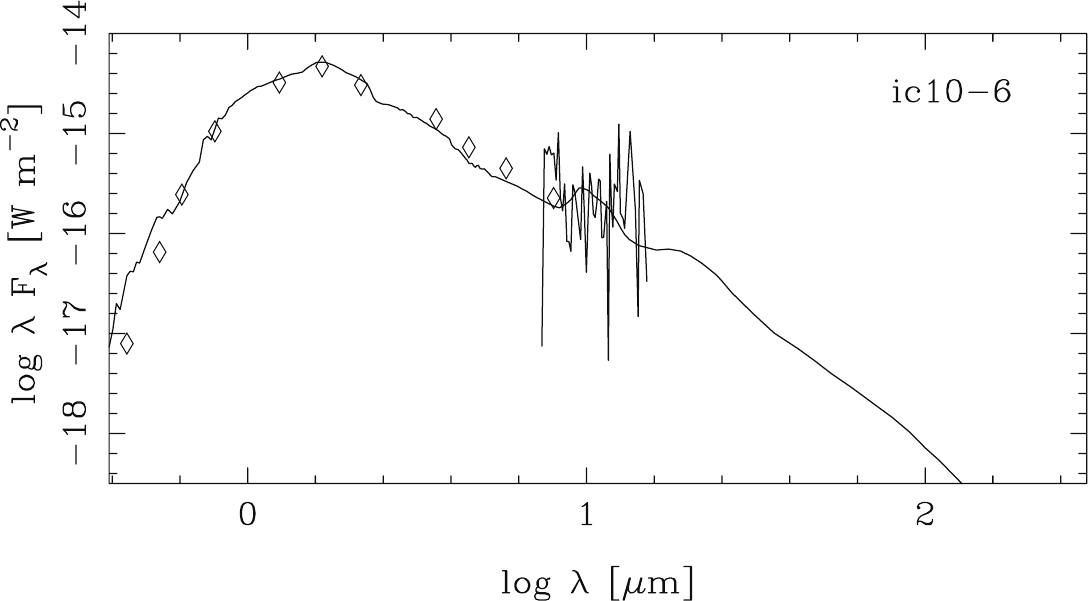}
\includegraphics[angle=0,scale=0.65,clip=true]{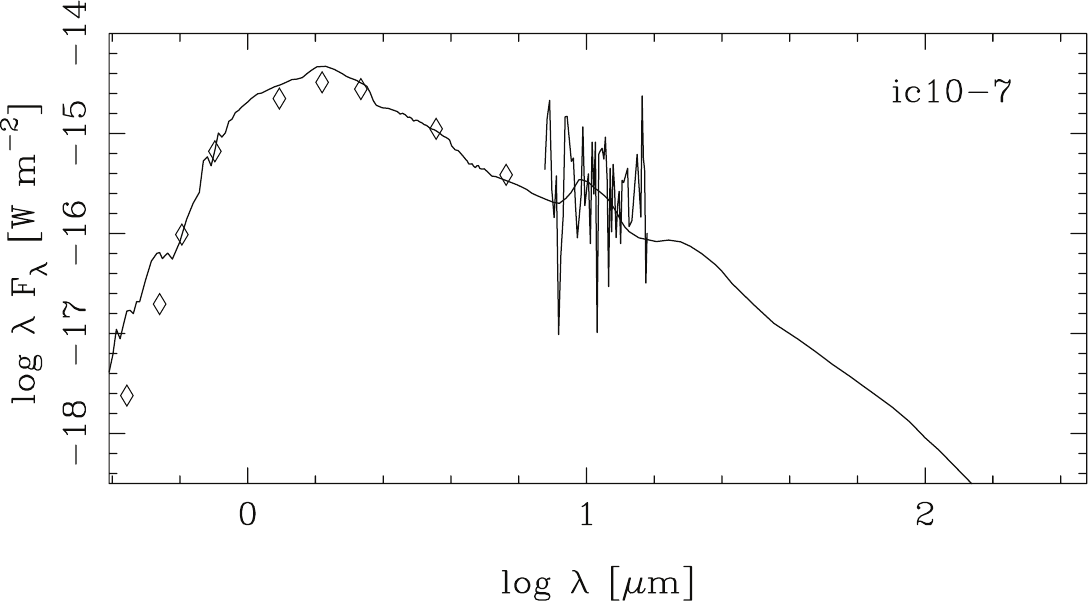}\\
\includegraphics[angle=0,scale=0.65,clip=true]{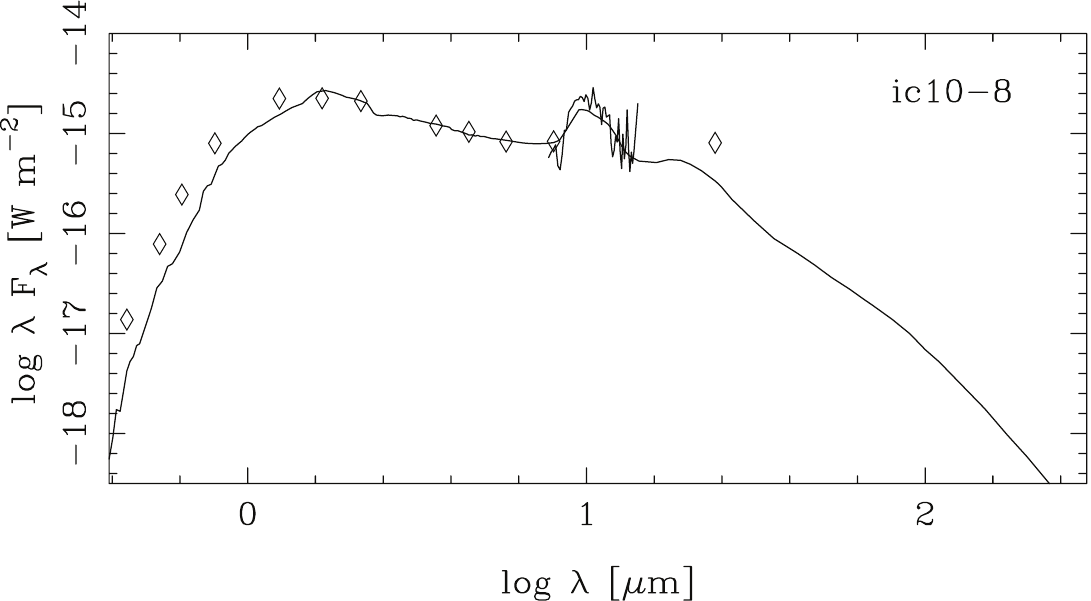}
\includegraphics[angle=0,scale=0.65,clip=true]{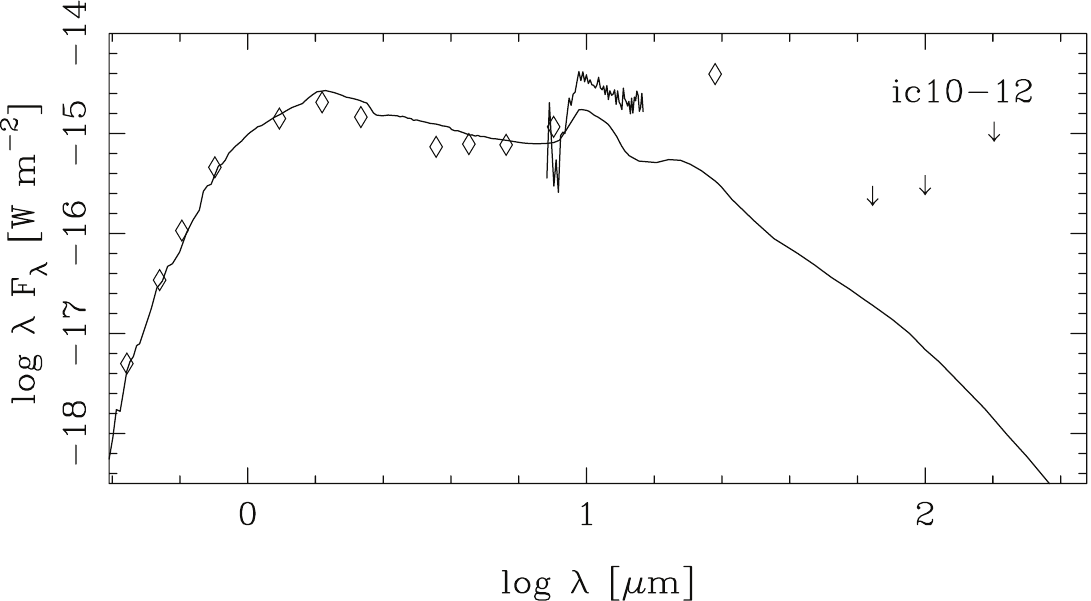}\\
\includegraphics[angle=0,scale=0.65,clip=true]{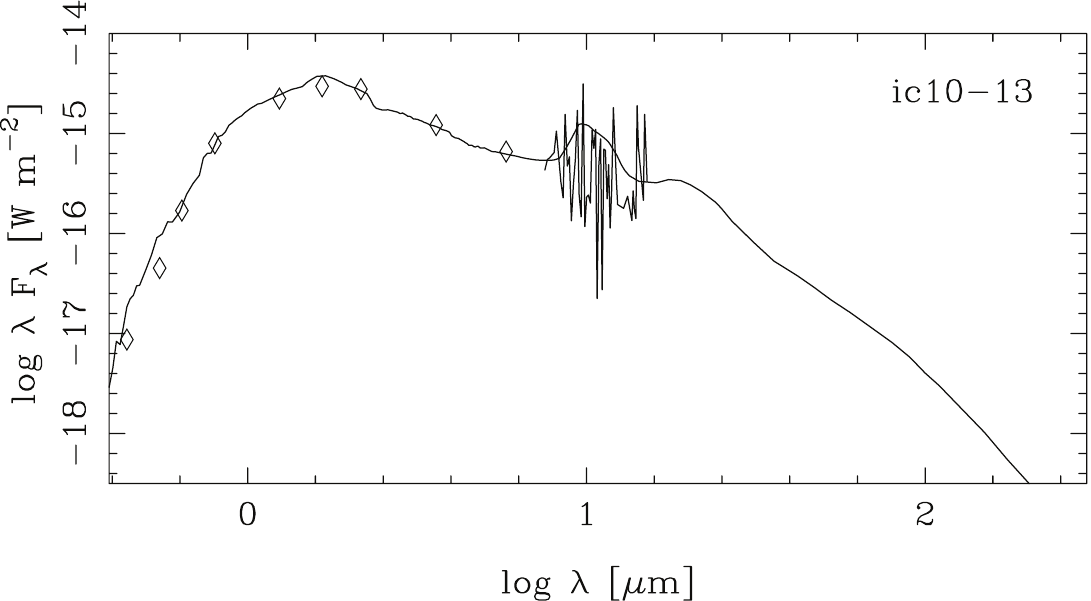}
\includegraphics[angle=0,scale=0.65,clip=true]{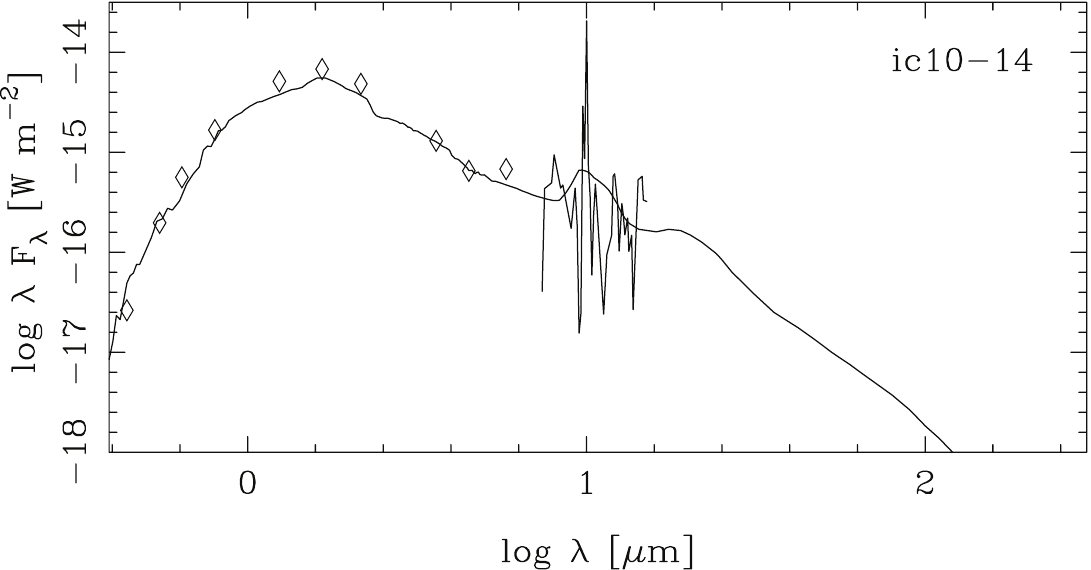}\\
\includegraphics[angle=0,scale=0.65,clip=true]{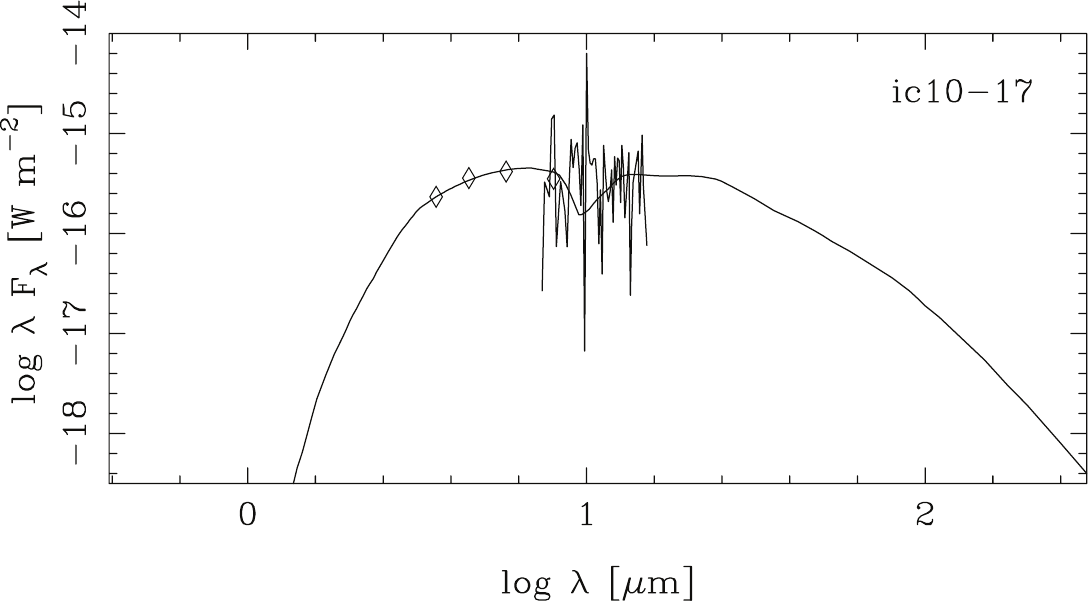}
  \caption{SED fit results. For each source, the SED model is shown on top as the solid curve, with the diamonds indicating photometry points from the optical bands, 2MASS, and Spitzer, and with the segments indicating the IRS data. The [24] data point for source \#8 bears significant uncertainties (see Table\,\ref{tab:irac}). The downward arrows for source \#12 show the upper limits on the Herschel/PACS fluxes. 
Source \#12 could not be fitted by our models (Section\,\ref{sec:model}). }
  \label{fig:models1}
  \end{center}
\end{figure*}

The most important constraint in distinguishing between AGBs and RSGs is the bolometric luminosity. It is expected that RSGs have a bolometric luminosity $M_{\rm bol}\lesssim-7.9$ (i.e., $\gtrsim$117\,000\,L$_\odot$), while AGBs should have $M_{\rm bol}\gtrsim-7.1$ ($\lesssim$56\,000\,L$_\odot$) (e.g., Wood et al.\ 1983).
Although AGBs can undergo hot bottom burning or thermal pulses that can increase their brightness temporarily (e.g., Groenewegen et al.\ 2009), the threshold $M_{\rm bol}\lesssim-7.9$ allows separating less luminous RSGs from intermediate-mass AGBs (see Massey et al.\ 2003; Massey \& Olsen 2003). 
We used the MIR spectra alongside the photometry to constrain the luminosity and mass-loss rate of the sources using the radiative transfer model described by Groenewegen et al.\ (1995, 2009).
For all stars we fitted a model with pure silicate dust (with absorption coefficients from Volk \& Kwok 1988), and another one with a mixture of 20\%\ aluminum oxide and 80\%\ silicate. For \#8 and \#12 the pure silicate dust model provided the best fit, while either model fits the data for the other sources. Table\,\ref{tab:ssil} and Figure\,\ref{fig:models1} shows the results for the silicate dust model.  
Based on the model results, most of the sources are much too bright to be AGBs. Only sources \#12 and \#17 fall below the luminosity threshold. Sources \#3, \#4, \#6, \#7, \#8, \#13, and \#14 are thus again compatible with RSGs.
%\textbf{In Kastner et al. 2008: JHK8 diagnostics plots (8um from MSX). Source 4 lies with RSGs, the others have K-A too large (assuming A = IRAC/8um...).}

%The specific case of source \#12 is puzzling, as it shows both strong silicate emission and a bright dust continuum longward of 10\mic\ visible from the IRS spectrum and from the MIPS 24\mic\ image. The SED of this object is not compatible with an RSG nature (Figure\,\ref{fig:models2}). 
%%This is confirmed by the fact that source \#12 was identified as a carbon star, [DBL2004]\,243, by Demers et al.\ (2004) using $R-I$ and CN$-$TiO colors. No, that source is 5\arcsec away. The SED cannot be reproduced by YSO models either (Robitaille et al.\ 2006; 2007), because it is far too luminous to be modeled by a single object. 
%On the other hand, it is possible to source \#12 is a very compact young cluster. The near-IR part of the SED could then arise from the main-sequence stars while the long-wavelength continuum would arise from dust heated by a young stellar generation. The optical and IRAC images show that this source is located in a crowded region, with several objects within a few arcseconds. It becomes a point-like source in IRAC [8.0\mic] and in the IRS wavelength range. No significant emission is detected above 70\mic\ with Herschel/PACS and SPIRE (S.\ Madden, private communication). %Preliminary upper limits at $3\sigma$: $5.6$\,mJy, $10.2$\,mJy, and $56.1$\,mJy at $70$\mic, $100$\mic, and $160$\mic\ respectively.

The specific case of source \#12 is puzzling, as it shows strong silicate emission, a dust continuum longwards of $10$\mic\ which is visible in the IRS spectrum, and it is detected in the MIPS $24$\mic\ image. The Herschel/PACS observations (S.\ Madden, private communication) place strict upper limits on the dust emission at far-infrared wavelengths. Due to the presence of a dust continuum, we compared the spectrum of this source to young star cluster dust radiative transfer models (Whelan et al.\ 2011) in order to test the hypothesis that source \#12 is a young compact stellar cluster. 
%The optical and IRAC images show that this source is located in a crowded region, with several objects within a few arcseconds. It becomes a point-like source in IRAC [8.0\mic] and in the IRS wavelength range.
If source \#12 is a young stellar cluster, then the near-IR measurements could arise from the main-sequence stars in the cluster while the long-wavelength continuum would arise from dust heated in the intra-cluster medium by the young stars. However, no models fit the data points: the $24$\mic\ data point and the PACS upper limits were much lower than expected for a range of appropriate dust geometries, from optically thin and geometrically thick to optically thin and geometrically thin. This suggests the lack of a carbonaceous grain dust component as one would expect in a dusty star forming environment. This finding is compatible with the lack of a clustering around source \#12 in the HST images (Figure\,\ref{fig:hst}). We therefore tentatively conclude that source \#12 is not a young compact cluster but is a dust-producing evolved star of some kind. Due to the presence of a silicate emission feature, low dust continuum, and relatively low luminosity (compared to RSGs) of planetary nebula NGC\,6804, source \#12 may be a planetary nebula around an O-star (Bilikova et al.\ 2012; Weidmann \& Gamen 2011), or a dusty WR star. The shallow long-wavelength SED is also reminiscent of extreme-AGB stars, which are usually carbon stars (see Boyer et al.\ 2012 and references therein). 

%3.5\arcsec at 660kpc is 1.69684788388e-05 rad = 11pc

%One source from our sample is identified as a carbon star in Demers et al.\ (2004): source \#3 as [DBL2004]\,162. Unfortunately, the detection of silicate dust is only tentative due to the low S/N. 

\subsection{Mass-loss rates}\label{sec:mlr}

%http://www.astro.keele.ac.uk/~jacco/research/vienna2006.pdf

Although it is possible to infer the mass-loss rate from molecular infrared transitions (Matsuura et al.\ 2006), the dust emission associated with the circumstellar envelope is the best tracer of mass loss. The mass-loss efficiency in O-rich stars depends more on metallicity than in C-rich stars, because O-rich dust depends on metallicity-limited elements (Si, Al), while amorphous carbon depends on self-produced C (e.g., Sloan et al.\ 2008, 2012). 

The dust production rate was computed from our models (Section\,\ref{sec:model}), assuming an outflow velocity of $10$\,km\,s$^{-1}$. The total mass-loss rate is estimated using a standard gas-to-dust ratio of 200.
The mass-loss rates we infer (Table\,\ref{tab:ssil}) lie within the ranges of what is found for the RSGs in the SMC and LMC, with rates between $10^{-5.5}$\,M$_\odot$\,yr$^{-1}$ and $10^{-7}$\,M$_\odot$\,yr$^{-1}$ (Groenewegen et al.\ 2009, assuming identical values for the outflow velocities and gas-to-dust ratio).

\subsection{Spatial distribution and sample completeness}\label{sec:similar}

A comparison of the luminosities of the RSG candidates (Table\,\ref{tab:ssil}) and theoretical isochrones (Fagotto et al.\ 1994) suggests that the stars in our sample (all but \#17) are more massive than $\gtrsim12$\,M$_\odot$. Their expected lifetime is $\sim20$\,Myr old, which is consistent with a starburst population. Is the spatial distribution of the RSG candidates compatible with the starbursting region? The claim for a starburst in IC\,10 mainly originates from the discovery of over 100 WR stars (Massey et al.\ 1992; Royer et al.\ 2001; Massey \&\ Holmes 2002). These studies showed that the spatial distribution of the WR stars is quite uniform, suggesting a widespread starburst. Hence, we do not expect to find the RSGs in any particular region, which is supported by our results.

We expect the number of RSGs to dominate over the number of WR stars at low metallicity, with a lower limit of $50$ RSGs in IC\,10 (e.g., Maeder et al.\ 1980; Massey et al.\ 2002, 2003). Our current sample sets a lower limit on the actual number of O-rich dust enshrouded stars and RSG candidates, with notably a strong limitation by the Spitzer/IRS sensitivity at 10\mic. 
In order to quantify the completeness of the sample, we built a sample of stars with similar colors as the RSG candidates we already identified.
Based on Sections\,\ref{sec:fore} and \ref{sec:model}, we choose the following constraints: $J-K=1.5\pm0.2$, $H-K>0.3$, $B-V>2.2$, and $V-R>1.6$. Only $16$ sources in the 2MASS/IRAC cross-matched sample (633 stars) fit these constraints, including the already confirmed sources \#4, \#8, and \#12 (Table\,\ref{tab:similar}). None of the other $13$ sources show silicate emission in their IRS spectra.
%, confirming that our study successfully detected the only silicate emission sources toward IC\,10. 
We partly attribute this low number of silicate emission sources to the low signal-to-noise ratio in the IRS spectra, as indicated by the IRAC [8.0] magnitudes. Only one source, 2MASS\,00200459+5918198, is expected to be bright enough for the silicate emission to be detected, and its [5.8]-[8.0] color does suggest the possible presence of silicate emission. However, the [8.0] flux is likely overestimated due to a contamination by PAH emission which might not have been completely subtracted when performing the aperture photometry with a sky annulus. A total of $13$ more sources could thus also be RSGs in addition to the $3$ sources we already identified. Considering an average mass-loss rate of $5\times10^{-7}$\,M$_\odot$\,yr$^{-1}$ (Section\,\ref{sec:mlr}), this results into a total mass-loss rate of $8\times10^{-6}$\,M$_\odot$\,yr$^{-1}$ for IC\,10 (see discussion in Section\,\ref{sec:discussion}). 

The low number of RSGs was already noticed by Massey et al.\ (2007), using deep optical images (photometry errors of $0.004$ in $V$ and $0.015$ in $B$ for $B=24.3$) and better spatial resolution than 2MASS and Spitzer observations. The authors argue that a very young burst ($\lesssim10$\,Myr) could be responsible for the large WR/RSG population ratio. Although our Spitzer data uncover just the tip of the RSG iceberg, the missing RSGs in IC\,10 remains an unsolved mystery.

\begin{table}
\begin{center}
  \caption{Sources with similar colors as confirmed silicate emission sources.}
  \label{tab:similar}
  \begin{tabular}{l l l l}
  \hline
2MASS source &   $K$  &  [8.0] & [5.8]-[8.0]\\
  \hline
00201538+5919070 & $13.52$ &$...$ & $...$ \\ 
00200259+5917481 (\#4) & $13.57$ &  $12.47$ & $0.36$ \\ 
00200322+5918013 & $13.67$ & $12.55$ & $0.27$ \\ 
00202465+5919003 & $13.92$ & $..$ & $...$ \\ 
00203019+5917154 & $14$ &  $...$ & $...$ \\ 
00200459+5918198 & $14.09$ &  $12.45$ & $0.78$ \\ 
00200510+5918039 (\#8) & $14.13$ &  $11.13$ & $1.02$ \\ 
00200825+5919092 & $14.26$ &  $...$ & $...$ \\ 
00200277+5917564 & $14.28$ & $13.21$ & $0.21$ \\ 
00202179+5917477 & $14.36$ &  $...$ & $...$ \\ 
00195375+5918118 & $14.43$ &  $...$ & $...$ \\ 
00202036+5918205 & $14.51$ &  $...$ & $...$ \\ 
00200839+5916419 & $14.51$ &  $14.07$ & $0.41$ \\ 
00201270+5917121 (\#12) & $14.51$ &  $10.77$ & $1.46$ \\ 
00200848+5916552 & $14.76$ &  $13.65$ & $0.37$ \\ 
00200819+5919202 & $15.06$ &  $...$ & $...$ \\ 
 \hline
  \end{tabular}
\tablefoot{Constraints on the colors are $J-K=1.5\pm0.2$, $H-K>0.3$, $B-V>2.2$, and $V-R>1.6$.}
%\tablenotetext{b}{Source not found by SExtractor.}
%\tablenotetext{b}{The distance between the centroid in the map and the 2MASS coordinates is shown between parentheses.}
\end{center}
\end{table}
%
%\subsection{Dust injection into the ISM}\label{sec:dust}
%
%Estimating the mass-loss rate of AGBs and RSGs is essential to apprehend dust production in the early Universe. AGB stars cannot be ruled out as significant contributors if one considers that the time to evolve from the zero-age main sequence (ZAMS) to the AGB phase depends on the metallicity. Ventura et al.\ (2002) and Herwig et al.\ (2004) find that AGBs can appear as early as 100\,Myr after the ZAMS below 1/30 solar metallicity. 

%http://web.ipac.caltech.edu/staff/ismevol/proceedings/zijlstraa_v1.pdf

\subsection{Discussion}\label{sec:discussion}

Our  analysis  of  stellar  properties  shows  that  the  luminosities  of  red-supergiants  in IC\,10 are  above  $90,000$\,L$_\odot$ (Table \,7).  	
Stellar  evolution  models show  that  stars  with  an  initial  mass  higher  than  $11.7$\,M$_\odot$ can  reach  luminosities  higher  than  $90,000$\,L$_\odot$  during  the  RSG phase (Fagotto  et  al.\  1994).  Stars  lower  than  $9$\,M$_\odot$  mass  cannot  reach  such  a  high  luminosity,  though  the  calculated  mass  range  lacks  $9-11.7$\,M$_\odot$ stars,  which  evolve  into  the  super-AGB  phase.  The  models  further  predict  that  the  age  to  reach such high luminosity phase is about $20$\,Myrs old or younger. 

The  age  of  RSGs  is  consistent  with  the star-formation  history  of  this  galaxy.  Hunter (2001)  analyzed  stellar  clusters  in  the  galaxy,  and  showed  that  IC\,10  has  episodes  of  high  star-formation;  young  ($4-30$\,Myrs)  clusters,  presumably  formed  in  the  starbursts,  and  intermediate age ($450$\,Myrs) clusters. The age of RSGs corresponds to the starburst phase  of this galaxy and it is consistent with the large population of WR stars detected (Massey \& Holmes 2002).

Our  analysis  might  provide a unique  case  of  measuring  mass-loss  rates  of  such  young  red-supergiants.  RSGs  in  IC\,10  might  belong  to  young  starbursts  ($4-30$\,Myrs),  while  LMC RSGs  are  mostly  from  intermediate  age  clusters  (a  few  100s of Myr;  Elson  \&  Fall  1988; van Loon et al.\ 1999). The mass-loss rates of young RSGs  are typically $10^{-7} - 10^{-6}$\,M$_\odot$\,yr$^{-1}$, which are comparable to those observed in LMC  RSGs.  We  did  not  detect  mass-loss  rates  higher  than $10^{-6}$\,M$_\odot$\,yr$^{-1}$  in our sample of  RSGs  (if removing  \#17),  although  such  higher  mass-loss  rates  have  been  found  in  Galactic RSGs  (e.g.,  VY\ CMa of  $10^{-4}$\,M$_\odot$\,yr$^{-1}$;  Decin  et  al.\ 2006).  This might be due to the fact that our selection of RSGs is limited by the detection limit of  the 2MASS photometry, where RSGs with high mass-loss rates are faint at near-IR wavelengths. 

The current analysis of mass-loss rates shows that gas ejected from the RSG population into the  ISM  is  at  least  $10^{-5}$\,M$_\odot$\,yr$^{-1}$,  and  the dust  return  from  RSGs  is  at  least  $5\times10^{-8}$\,M$_\odot$\,yr$^{-1}$.  We consider this to be a lower limit, because it is likely that stars with high mass-loss rates are not  detected yet (because of dust extinction in the K band),  but  these  stars  could  to contribute  significant  fraction  to  gas  and  dust  return  from the total RSG populations into the ISM. Compared with the gas mass of IC\,10 ($\sim10^8$\,M$_\odot$, Yin et al.\ 2010) and a star-formation rate of  up to $0.2$\,M$_\odot$\,yr$^{-1}$ (Leroy  et  al.\ 2006),  which  represents the ISM  gas  mass  consumed  by  the  formation  of  stars,  the  mass  injected  from  RSGs  is  significantly  small,  and has little  impact  on  the  total  gas  ISM  mass  at  the  current  stage.  Although  the  total  mass  of  ISM  dust  in  this  galaxy  is  unknown, it is likely to be on the  order of $10^6$\,M$_\odot$, considering the gas-to-dust mass ratio.  The dust from RSGs appear to be not an important contributor to the dust mass in this galaxy if the star formation rate of this galaxy has been more or less similar for the past few Myrs. A similar conclusion was reached for the Magellanic Clouds where AGBs are shown to dominate the stellar dust production (e.g., Boyer et al.\ 2012). We conclude that although RSGs could in principle dominate the dust production over AGBs in a starbursting galaxy (e.g., Massey et al.\ 2005), this is not observed in IC\,10.

\section{Conclusions}

We report the discovery of O-rich dust enshrouded stars within the nearby ($\approx700$\,kpc) dwarf starburst galaxy IC\,10. We examined the Spitzer/IRS spectral map ($7.5-14.5$\mic) in order to build a sample of point-like sources potentially showing silicate dust in emission. The silicate strength map we constructed reveals several point-like sources and no extended emission. 

Most sources are associated with single, point-like, 2MASS and optical sources. We investigate the colors and magnitudes in the near-IR and optical, and identify $9$ sources belonging to the IC\,10 system. The colors and photometry in the optical and near-infrared suggests that these sources are different from AGB stars. Modeling of the dust predicts high luminosities compatible with RSGs. This is thus the farthest detection of O-rich dust encircled stars confirmed spectroscopically. The low number of sources discovered spectroscopically does not solve the apparent lack of RSGs as compared to WR stars in IC\,10 (Massey et al.\ 2007). 

We derived mass-loss rates for all sources using a radiative transfer model. Accounting for sample completeness, the total mass-loss rate is significantly small as compared to the dust mass in IC\,10. Other sources of dust (AGBs, SNe, WR) are necessary to explain the dust mass observed in the ISM of IC\,10. 

Another source belonging to the IC\,10 system shows strong silicate emission together with a warm carbonaceous dust grain continuum. The nature of this source remains unknown.

\begin{acknowledgements}
We thank the anonymous referee for a helpful report. This work is based on observations made with the Spitzer Space Telescope, which is operated by the Jet Propulsion Laboratory, California Institute of Technology under a contract with NASA.
This study made use of the Two Micron All Sky Survey (2MASS)
M.F. Skrutskie, R.M. Cutri, R. Stiening, M.D. Weinberg, S. Schneider, J.M. Carpenter, C. Beichman, R. Capps, T. Chester, J. Elias, J. Huchra, J. Liebert, C. Lonsdale, D.G. Monet, S. Price, P. Seitzer, T. Jarrett, J.D. Kirkpatrick, J. Gizis, E. Howard, T. Evans, J. Fowler, L. Fullmer, R. Hurt, R. Light, E.L. Kopan, K.A. Marsh, H.L. McCallon, R. Tam, S. Van Dyk, and S. Wheelock, 2006, AJ, 131, 1163. 
JBS wishes to acknowledge the support from a Marie Curie Intra-European Fellowship within the 7th European Community Framework Program under project number 272820. 
P.M.'s contributions to this project were supported by the National Science Foundation under grant AST-1008020.
\end{acknowledgements}

\appendix

\section{Cross-correlation LGGS stars $-$ silicate emission candidates}

The silicate emission candidates from Table\,\ref{tab:sources} are matched with optical sources from the LGGS in Section\,\ref{sec:lggs}. When several LGGS stars fall within $4\arcsec$ of the silicate emission source coordinates, we selected the closest LGGS star. We list in Tables\,\ref{tab:contam1} and \ref{tab:contam2} the LGGS sources within $4\arcsec$ from each silicate emission candidate. We also list their optical colors. In all cases, the closest LGGs star is also the best match in terms of brightness and red color. 

\begin{table}
\begin{center}
  \caption{LGGS sources within $d<4\arcsec$ of the silicate emission candidates identified as members of IC\,10.}
  \label{tab:contam2}
  \begin{tabular}{l l l l l l}
  \hline
  ID & $d$ ($\arcsec$) & LGGS  & $V$ & $V-R$ & $R-I$ \\
  \hline
\#3 & 2.33&	J002001.84+591933.9&	23.463&	0.684&	...  \\ %22.8 confusion there is really nothing in IRAC there, looks like this star is just not in LGGS
	&2.89&	J002001.22+591934.6&	22.928&	0.647&	1.032 \\ %21.3     confusion
	&3.40&	J002001.82+591930.5&	22.891&	0.691&	0.653 \\ %21.5
	&3.83&	J002001.96+591935.4&	23.475&	1.025&	0.904 \\ % 21.6
\hline
\#4 & 0.18&	\textbf{J002002.61+591748.2} &	21.629&	2.008&	2.192 \\ %17.4
	&2.22&	J002002.55+591750.3&	23.484&	1.020&	1.241\\ %21.3 less red
	&3.29&	J002002.94+591746.2&	21.804&	0.492&	0.480\\ %20.8 ;faint in IRAC ch1 
	&3.79&	J002003.05+591749.5&	22.522&	0.440&	... \\ %R=22.1
\hline
\#6 & 0.31&	\textbf{J002003.23+591801.6} &	21.274&	1.848&	2.035 \\ %17.3
	&2.27&	J002003.48+591800.2&	23.084&	0.537&	...  \\ %R=22.6
	&2.59&	J002003.31+591803.8&	23.142&	0.938&	1.183 \\ %21 less red
	&3.77&	J002002.94+591804.4&	21.656&	0.745&	0.848 \\ %20
	&3.91&	J002003.25+591805.2&	23.494&	0.703&	...  \\ %22.8
\hline
\#7 & 0.25&	\textbf{J002004.54+591852.3} &	22.559&	2.207&	2.428 \\ %18
\hline
\#8 & 0.21&	\textbf{J002005.11+591804.1} &	21.096&	1.666&	1.717 \\ %17.7
	&0.98&	J002005.19+591804.6&	21.529&	0.238&	1.206 \\ %20.1 %less red, less bright, within the IRAC ch1 emission though
	&2.52&	J002005.00+591806.3&	22.979&	0.720&	0.601 \\ %21.7
	&3.60&	J002005.45+591801.5&	19.672&	0.875&	0.997 \\ %17.8
\hline
\#12 & 0.30&	\textbf{J002012.73+591712.3} &	22.020&	1.722&	1.978 \\ %18.3 
	&2.23&	J002012.75+591709.9&	22.271&	1.535&	1.453 \\ %19.3 %less red less bright, XX (bright in IRAC ch1) not even sure which one is the one from the SiO map
	&2.51&	J002013.00+591713.1&	23.496&	0.845&	1.139 \\%21.5
	&3.02&	J002012.59+591715.0&	22.626&	0.689&	1.024 \\ %20.9
	&3.64&	J002012.90+591715.4&	23.185&	0.928&	1.220 \\ %21.1
	&3.82&	J002013.19+591711.4&	21.644&	0.909&	0.976 \\ %19.7
\hline
\#13 & 0.25&	\textbf{J002022.28+591743.3} &	21.682&	1.870&	2.095 \\ %17.7
	&1.54&	J002022.45+591743.0&	22.817&	0.632&	...  \\ %R=22.2 within the IRAC ch1 blob
	&2.49&	J002022.46+591741.3&	21.704&	0.358&	0.290 \\ %21
	&2.93&	J002021.97+591741.2&	20.790&	0.492&	0.141 \\ %20.2 maybe in IRAC ch1 ?
	&3.30&	J002022.23+591746.5&	22.346&	2.662&	1.090 \\ %18.6 ;bright in ch1
	&3.33&	J002022.13+591746.4&	20.599&	1.013&	0.972 \\ %18.6 ;bright in ch1
	&3.39&	J002022.68+591742.4&	21.850&	0.478&	0.779 \\ %20.6 ;faint in irac ch1
	&3.59&	J002021.87+591741.1&	19.094&	0.452&	0.447 \\ %18.2 ;faint in ch1
\hline 
\#14 & 0.18&	\textbf{J002022.01+591724.5} &	20.096&	1.581&	1.626 \\ %16.9 
	&0.84&	J002022.05+591725.1&	22.266&	3.631&	1.774 \\ %16.9 red, close ;in the ch1 blob
	&1.63&	J002022.10+591725.8&	21.842&	1.368&	0.601 \\ %19.8 ;in the ch1 blob
	&1.76&	J002022.11+591725.9&	22.051&	1.566&	0.616 \\ %19.9 ;in the ch1 blob
	&1.89&	J002021.78+591725.4&	22.936&	0.132&	...  \\ %R=22.8 ;in the ch1 blob
	&2.25&	J002022.22+591723.0&	23.046&	1.235&	1.264 \\ %20.5 ;in the ch1 blob
	&2.35&	J002022.05+591722.1&	21.650&	0.932&	1.078 \\ %19.7 ;?
	&2.45&	J002021.85+591722.2&	22.018&	0.846&	1.099 \\ %20 ;?
	&2.82&	J002021.82+591726.9&	23.105&	1.172&	1.418 \\ %20.5
	&3.75&	J002021.68+591721.5&	20.448&	0.514&	0.444 \\ %19.5
	&3.94&	J002022.44+591726.3&	22.516&	0.469&	0.702 \\%21.3
	&3.96&	J002021.60+591721.8&	20.446&	0.512&	0.444 \\ %19.6
\hline
  \end{tabular}
  \tablefoot{The LGGS ID in bold indicates the best match in terms of distance and infrared brightness. }
  \end{center}
\end{table}

\begin{table}
\begin{center}
  \caption{LGGS sources within $d<4\arcsec$ of the silicate emission candidates identified as field stars.}
  \label{tab:contam1}
  \begin{tabular}{l l l l l l}
  \hline
  ID & $d$ ($\arcsec$) & LGGS & $V$ & $V-R$ & $R-I$ \\
  \hline
\#2 & 0.80 & \textbf{J001957.61+591835.5} &	20.938 &	1.036&	1.132 \\ % I = 18.8
	&3.12&	J001957.83+591837.8&	23.890&	0.997&	1.084 \\ %  21.8
	&3.49&	J001957.24+591835.8&	22.182&	1.589&	1.717 \\ % 18.9 %far but very red
	&3.59&	J001957.99+591832.2&	22.946&	0.550&	0.604 \\ % 21.7
\hline
\#10 & 1.09&	\textbf{J002011.91+591827.6} &	19.911&	0.883&	0.955 \\ % 18; faint
	&2.51&	J002011.55+591825.4&	21.903&	0.672&	  \\ %R= 21.2 ;in ch1 blob
	&3.02&	J002011.87+591829.7&	22.798&	0.438&	  \\ %R=22.4
	&3.74&	J002011.76+591823.0&	22.901&	1.456&	1.408 \\% 20 %far but very red
	&3.95&	J002011.59+591823.2&	23.305&	1.091&	1.452 \\ %20.7%far but very red
\hline
\#11 & 3.78&	J002012.61+591724.6&	23.381&	0.834&	0.997 \\ %21.6
\hline
\#15 & 0.97&	\textbf{J002022.52+591732.9} &	18.404&	0.590&	0.642 \\ %17.2
	&1.81&	J002022.17+591732.8&	22.633&	1.026&	1.019 \\ %20.6  ;in ch1 blob
	&2.46&	J002022.33+591730.8&	22.744&	1.521&	1.460 \\ %19.7 %not so far, very red; ?
	&3.52&	J002021.99+591731.6&	23.073&	0.443&	0.921 \\ % 21.8
	&3.60&	J002022.40+591729.6&	23.417&	1.003&	0.782 \\ %21.6 ;?
	&3.82&	J002022.20+591729.7&	23.299&	-0.084&	\\ %R=23.4
\hline
\#16 & 0.38&	\textbf{J002025.23+591807.3} &	20.857&	1.409&	0.949 \\ %18.6 
	&0.46&	J002025.14+591807.0&	19.644&	0.263&	0.412 \\ %18.9 %close although less red ; ? in ch1
	&1.29&	J002025.36+591806.6&	20.557&	1.082&	1.644 \\ %17.9 %close, red  ;in the ch1 blob
	&1.40&	J002025.05+591807.8&	20.807&	0.459&	0.188 \\ %20.1 ;?
	&1.64&	J002025.41+591806.7&	21.316&	0.165&	  \\ %R=21.1 ;?
	&1.92&	J002025.24+591805.1&	19.392&	0.712&	0.724 \\ %18  ;in the ch1 blob
	&1.97&	J002025.33+591808.7&	21.954&	0.427&	  \\%R=21.5 ;?
	&2.04&	J002024.95+591806.3&	19.696&	0.437&	0.408 \\ %18.9 ;in the ch1 blob
	&3.14&	J002025.60+591807.7&	21.873&	0.433&	0.415 \\ %21.1 
	&3.39&	J002025.45+591804.2&	21.704&	1.354&	1.352 \\ %19 ;faint in ch1
	&3.43&	J002025.04+591810.2&	22.570&	-0.150&	  \\ %R=22.7
	&3.48&	J002025.47+591804.2&	21.753&	1.400&	1.355  \\ %19.1 ;faint in ch1
\hline
  \end{tabular} 
   \tablefoot{The LGGS ID in bold indicates the best match in terms of distance and infrared brightness. }
\end{center}
\end{table}

\end{document}